\documentclass[prl,reprint,aps,superscriptaddress,longbibliography,floatfix]{revtex4-2}
\usepackage{graphicx}
\usepackage{bm}
\usepackage{amsmath}
\usepackage{amssymb}
\usepackage{booktabs}
\usepackage{multirow}
\usepackage{enumerate}
\usepackage{mhchem}
\usepackage{upgreek}
\usepackage[dvipsnames]{xcolor}
\usepackage{comment}

\usepackage[normalem]{ulem} 
\usepackage[hidelinks]{hyperref}
\hypersetup{colorlinks=true, linkcolor=blue, citecolor=red, urlcolor=blue}

\newcommand{\lp}{\left}
\newcommand{\rp}{\right}

\newcommand{\Mean}[1]{\lp\langle{#1}\rp\rangle}

\newcommand{\0}{{\bm 0}}
\newcommand{\dd}{{\bm d}}

\newcommand{\GG}{{\bm G}}

\newcommand{\rr}{{\bm r}}

\newcommand{\DDelta}{{\bm\Delta}}

\newcommand{\ddelta}{{\bm\delta}}

\newcommand{\ttau}{{\bm\tau}}
\newcommand{\pphi}{{\bm\phi}}
\newcommand{\nnabla}{{\bm\nabla}}
\newcommand{\Lambdabar}{{\overline\Lambda}}

\newcommand{\mcE}{{\mathcal{E}}}

\newcommand{\mcP}{{\mathcal{P}}}
\newcommand{\mcR}{{\mathcal{R}}}

\renewcommand{\Re}{{\mathrm{Re}}}
\renewcommand{\Im}{{\mathrm{Im}}}

\begin{document}

\title{Spontaneous Twirls and Structural Frustration in Moir\'e Materials}

\author{Jingtian Shi}
\affiliation{Materials Science Division, Argonne National Laboratory, Lemont, Illinois 60439, USA}

\author{Gaurav Chaudhary}
\affiliation{TCM Group, Cavendish Laboratory, University of Cambridge, Cambridge CB3 0US, United Kingdom}

\author{Allan H. MacDonald}
\affiliation{Department of Physics, University of Texas at Austin, Austin TX 78712, USA}

\author{Ivar Martin}
\affiliation{Materials Science Division, Argonne National Laboratory, Lemont, Illinois 60439, USA}

\date{\today}

\begin{abstract}

Structural twirls form spontaneously in the domain wall networks of some moir\'e materials. 
We show that in heterobilayers, neighboring twirl chiralities tend to anti-align, forming staggered patterns that
are well described by antiferromagnetic lattice $\phi^4$ theories. In moir\'e systems with triangular domains, this leads to frustration in the chirality 
configuration of the structural twirls and to hysteresis with respect to variation of the average twist angle and possibly other control parameters.

\end{abstract}

\maketitle

\paragraph{Introduction.}
Moir\'e superlattices arise from interference between stacked layers of two-dimensional (2D) materials with small lattice-constant mismatches and/or relative twists.  
When the host materials are semiconductors or semimetals, they behave as  
artificial two-dimensional solids -- moir\'e materials --
with lattice constants on $\sim 10\,\rm nm$ length scale.
The large lattice constants allow band filling factors to be tuned 
through large intervals with electrical gates, a property that has been key to
the realization of a wide range of exotic quantum states in moir\'e materials \cite{cao2018correlated, cao2018unconventional, cai2023signatures,andrei2021marvels}. 
Individual layers in moir\'e materials are known to undergo lattice-strain 
relaxation that expands regions with energetically favorable local stacking arrangements, and in the limit of large moir\'e periods shrinks regions with unfavorable stacking arrangements into domain walls \cite{carr2018relaxation}.
Domain structures of this type have been extensively observed across moir\'e systems
using a variety of techniques \cite{alden2013strain,yoo2019atomic,weston2020atomic,molino2023ferroelectric,Mcgilly2020visualization}.
\begin{figure}[t]
    \centering
    \includegraphics[width=0.48\textwidth]{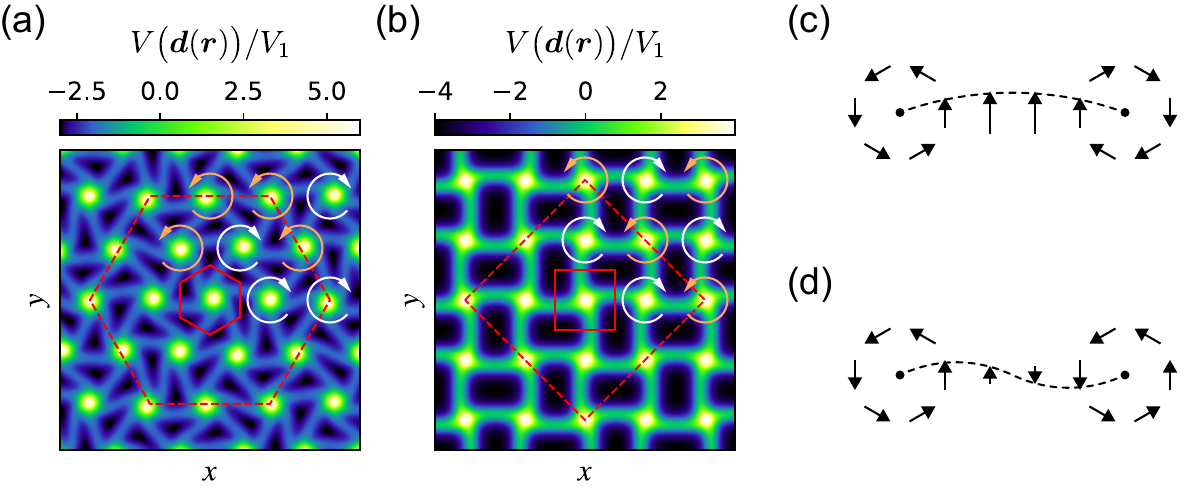}
    \caption{(a)-(b) Typical stacking energy density map of relaxed aligned heterobilayer moir\'es with lattice mismatch $\epsilon = -0.017$ and $V_1 = 3\times10^{-4}\mu$ (see definitions in the text): (a) honeycomb lattice with triangular domains, \textit{i.e.}, the triangular system; (b) square lattice. The curved arrows mark the spontaneous twirls of domain wall junctions. The solid and dashed red hexagons or squares are respectively the original moir\'e cell and the supercell periodicity imposed in our calculations. (c)-(d) Schematics of neighboring twirls with (c) opposite chiralities and (d) the same chirality. The small arrows represent the local atomic displacement, and the dashed lines sketch the trend of shear displacement.}
    \label{fig:introduction}
\end{figure}
In some of these systems, an additional intriguing behavior has been predicted \cite{dai2016twisted, quan2018tunable, maity2021reconstruction, mesple2023giant, kaliteevski2023twirling}
-- namely that the domain wall networks can spontaneously develop twirl distortions. Recent studies \cite{dejong2023stacking, mesple2023giant, jin2020van} have provided partial experimental confirmation.  The twirling induces spontaneous 
breaking of symmetries, including moir\'e translational symmetries. This process reduces total energy by converting a part of tensile/compression energy into less costly shear strain \cite{mesple2023giant, kaliteevski2023twirling, soltero2024competition}.
In this work, we show that this process can also lead to a novel form of structural frustration.
Fig. \ref{fig:introduction} (a) shows an example of a distorted triangular domain wall network we obtained from  continuum elasticity model
\cite{san-jose2014spontaneous, nam2017lattice, bennett2022theory} of lattice-mismatched heterobilayer honeycomb lattices, in which we see an irregularly staggered arrangement of twirls at domain wall junctions. By comparison, a checkerboard pattern is observed in the domain wall network of a heterobilayer square lattice, as shown in Fig. \ref{fig:introduction} (b).
These phenomena reveal an effective ``{\em antiferromagnetic}'' coupling
between the chiralities of neighboring twirls, which is frustrated in triangular lattices and leads to a N\'eel-like checkerboard pattern 
The tendency to anti-align can be intuitively understood from Figs. \ref{fig:introduction} (c), (d): for anti-aligned (aligned) neighboring twirls, there is a smaller (larger) shear strain in the mid-region between the twirls.

While the staggered arrangements of twirl chiralities is suggestive of the possible description of the system in terms of a classical antiferromagnetic Ising model, we show that the typical configurations and their energies are more accurately described by an antiferromagnetic lattice $\phi^4$ model, in which the discrete spins of the Ising model are replaced by continuous real scalar degrees of freedom that have double-well onsite potentials. In this way, the  variations in the intensities of the twirls are captured, which is beyond the simple binary description in terms of the twirl chirality. Similar to a spin glass, structural frustration in the triangular domain wall network leads to exponentially many nearly degenerate configurations. We show that this glassiness can manifest in hysteretic behavior of twirl arrangement in response to variation
of twist angles and other control parameters.

\smallskip
\paragraph{Model.}
We base our calculations on the effective model of a single layer of an elastic 2D material on top of a rigid substrate~\cite{san-jose2014spontaneous}. This model is also sufficient to describe the lattice relaxation of bilayers consisting of two elastic 2D materials with equal Poisson ratios \cite{nam2017lattice, koshino2019moire, supplemental}.
For a generic moir\'e superlattice, we consider the elastic material layer rotated counter-clockwise by a twist angle $\theta$ relative to the substrate and a lattice constant ratio of $1+\epsilon$ between the material and the substrate.
We define the local registry parameter $\dd(\rr)$ as the local displacement of the material lattice with respect to the substrate lattice.
In the rigid case, the local registry parameter is 
\begin{equation}
    \label{eq:registryrigid}
\dd_0(\rr) = \lp((1+\epsilon)\mcR_\theta - 1\rp)\rr,
\end{equation}
where $\mcR_\theta$ denotes the $2\times 2$ rotation matrix 
for angle $\theta$. Accordingly, the moir\'e superlattice is related to the Bravais lattice of the material via the transformation matrix $\lp((1+\epsilon)\mcR_\theta - 1\rp)^{-1}$.

We adopt the standard 2-dimensional continuum elasticity model \cite{san-jose2014spontaneous, nam2017lattice, bennett2022theory} to describe the lattice relaxation.
For simplicity, we limit our degrees of freedom to the in-plane displacement,
denoting the displacement from the rigid position as $\ddelta(\rr)$, such that $\dd(\rr) = \dd_0(\rr) + \ddelta(\rr)$. The equilibrium configuration can be obtained by minimizing the total lattice energy
\begin{equation}
	E = \int d^2\rr  \; (U + V),
    \label{eq:totalenergy}
\end{equation}
where the elastic energy density $U$ is
\begin{equation}
	U = \frac{\lambda}{2} (\nnabla\cdot\ddelta)^2 + \frac{\mu}{4} \sum_{\alpha, \beta = x, y}(\partial_\alpha\delta_\beta + \partial_\beta\delta_\alpha)^2
    \label{eq:elasticenergy}.
\end{equation}
Here $\lambda$ and $\mu$ are the Lam\'e parameters of the monolayer material, and the stacking energy density $V$ is a periodic function in the registry vector $\dd$ with the periodicity of the material's lattice. Therefore,
\begin{equation}
	V(\dd) = \sum_{\GG\in\Lambdabar} V_\GG e^{i\GG\cdot\dd},
    \label{eq:V(d)}
\end{equation}
where $\Lambdabar$ is the reciprocal lattice of the material. 
(Vertical relaxation effects are 
assumed to be also incorporated in $V$.)
We truncate $V_\GG$ into the first star of the reciprocal lattice, \textit{i.e.} $\Lambdabar_1 = \{\GG_0, \GG_1, \dots, \GG_{m-1}\}$, where $\GG_j = G(-\sin(2j\pi/m), \cos(2j\pi/m))$, $m = 6$ ($4$), $G = 4\pi/\sqrt{3}a$ ($2\pi/a$) for triangular (square) systems
\footnote{Throughout this paper we refer to the honeycomb lattice moir\'e with triangular-lattice arrangement of domain wall junctions as the ``triangular system''.}, and $a$ is the lattice constant.
All $V_{\GG_j}$'s are taken to be equal to a real positive parameter $V_1$ \footnote{Negative $V_1$ leads to hexagonal domains, where the domain wall junctions are arranged in hexagonal rather than triangular form. Frustration does not occur in this case.
In square lattices, $V_1$ can be converted to $-V_1$ by a redefinition $\dd \rightarrow \dd + (a/2, \,a/2)$.} to capture the
6- (4-) fold rotational symmetry of the triangular (square) system.
Our model setting for the triangular system approximates transition metal dichalcogenide (TMD) bilayers, although small imaginary parts of $V_{\GG_j}$ are expected to be present in those cases due to the distinction of A and B sublattices \cite{bennett2022theory}. Non-zero $\Im V_{\GG_j}$ values reduce the 6-fold symmetry to 3-fold symmetry, and their influence is discussed in the Supplemental Material (SM) \cite{supplemental}.

Using the variational condition $\delta E/\delta\ddelta = \0$, we obtain a nonlinear differential equation for $\ddelta$,
\begin{equation}
	\mu\nnabla^2\ddelta + (\lambda+\mu) \nnabla (\nnabla\cdot\ddelta) = iV_1\sum_{j = 0}^{m-1} \GG_j e^{i\GG_j\cdot [\dd_0(\rr) + \ddelta(\rr)]}.
    \label{eq:relax}
\end{equation}
We see that the order of magnitude of $\ddelta$ is roughly $\sim V_1|\GG_j|a_M^2/\mu \sim aV_1/\mu\epsilon^2$, where $a_M$ is the  moir\'e  lattice scale, and $a$ is the microscopic lattice constant. 
If $\ddelta$ on the right-hand side can be neglected, then the symmetry of the solution should match the symmetry of the unrelaxed moir\'e pattern. However, nonlinearity can stabilize a symmetry-broken solutions. That will happen when $\ddelta$ in the exponential becomes of order $a$, \textit{i.e.}, 
$V_1|\GG_j|^2 a_M^2/\mu \sim V_1/\mu \epsilon^2\sim 1$.
In typical materials $V_1/\mu \sim 10^{-4}$, which means that for $\epsilon \lesssim 10^{-2}$ a nonlinear regime is reached. 
We numerically solve Eq. (\ref{eq:relax}) starting from random  initial conditions for  $\ddelta(\rr)$.  In each step, we represent the right-hand side on a $36\times36$ ($40\times40$) real-space grid in each moir\'e cell of the triangular (square) system and solve for $\ddelta$ on the left-hand side by expansion in Fourier space. In addition, in the triangular (square) system we impose periodic boundary condition in $\ddelta(\rr)$ 
over a supercell that is a $2\sqrt{3} \times 2\sqrt{3}$ ($2\sqrt{2} \times 2\sqrt{2}$) expansion of the moir\'e unit cell. 
This enlargement allows for ground state configurations that break the 
moir\'e translational symmetry.  Below we focus mainly on 
the triangular system.

\smallskip
\paragraph{Analysis of Twirl Configurations.}
First, we focus on aligned heterobilayers ($\theta = 0$) and present calculations with a lattice constant mismatch $\epsilon = -0.017$ and a Lam\'e constant ratio $\lambda/\mu = 0.5$. Our choice of $\epsilon$ corresponds to the lattice mismatch at a heterobilayer between graphene and hexagonal boron nitride (hBN) \cite{haastrup2018computational, gjerding2021recent}, and the Lam\'e constant ratio is in the typical range for
2D materials including graphene, hBN and TMDs \cite{sachs2011adhesion, carr2018relaxation}. 
We further choose $V_1/\mu = 3\times 10^{-4}$ -- strong enough to create twirls.
Results for weaker $V_1$ are presented in the SM \cite{supplemental}.
Strikingly, for the triangular system, we find many metastable configurations with very similar total energies. Each has irregularly arranged left (counterclockwise) and right (clockwise) twirls at the domain wall junctions, leading to a complex distorted domain wall network. 
Fig. \ref{fig:introduction}(a) illustrates the lowest-energy example of these configurations by plotting its local stacking energy density $V(\dd(\rr))$ {\it vs.} position $\rr$.
This is in contrast with the square lattice result under the same parameter choice, as shown in Fig. \ref{fig:introduction}(b), which is the only stable configuration (up to universal translations).
As argued in the introduction, the metastable states of the triangular system are results of frustration.

\begin{figure}
    \centering
    \includegraphics[width=0.48\textwidth]{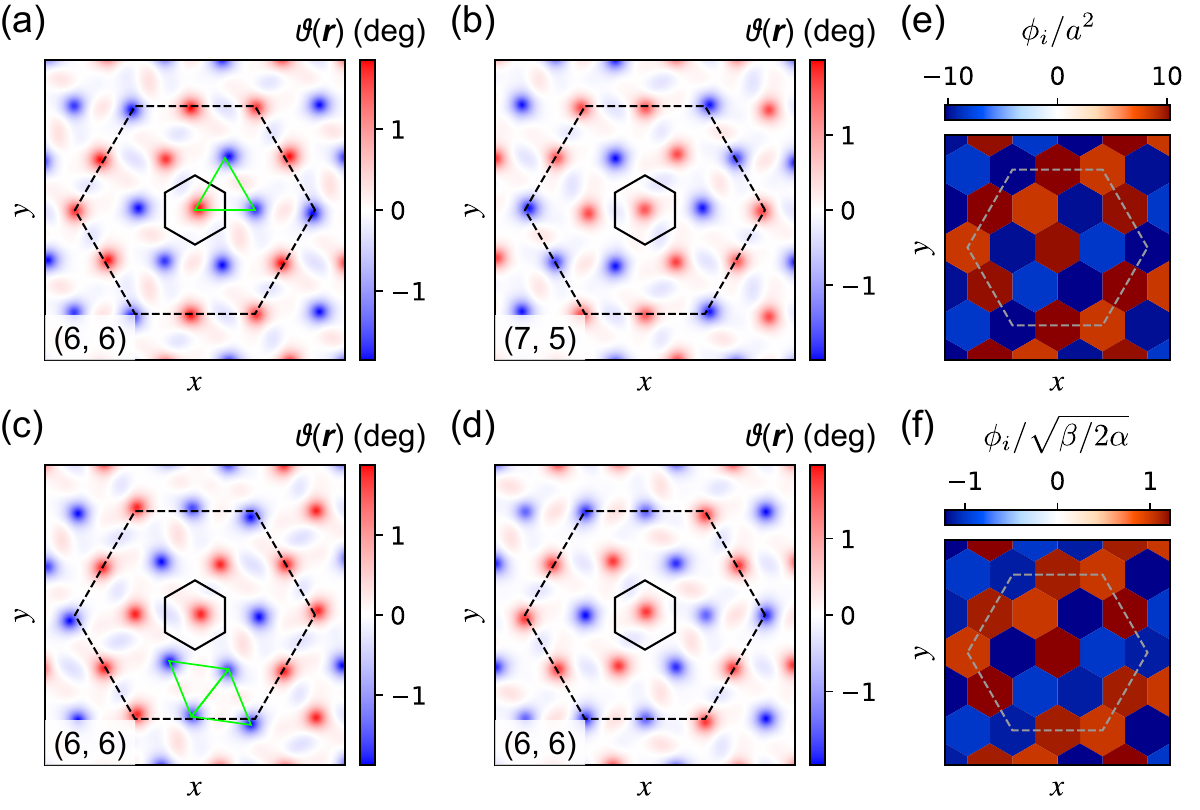}
    \caption{(a)-(d): Local twist angle plots for the four lowest 
    energy metastable configurations of the triangular system. ((a)-(d) are 
    in ascending order of total energy.) In these plots we have taken $\epsilon = -0.017$, $\lambda = \mu/2$ and $V_1 = 3\times 10^{-4}\mu$. In each panel, the numbers of left and right twirls in each $2\sqrt{3} \times 2\sqrt{3}$ supercell are marked at the bottom left. The green triangle in (a) is an example of a frustrated triangular plaquette.
    The green lines in (c) outline a region in which  the number of 
    frustrated links is not minimized.
    (e) The twirl moment (see main text) configuration of the lowest-energy state plotted in (a). The color of each patch corresponds to the total twirl moment $\phi_i$ in the corresponding moir\'e Wigner-Seitz cell. (f) The lowest-energy configuration of the nearest-neighbor antiferromagnetic lattice $\phi^4$ model
    (Eq. (\ref{eq:AFM_phi4})) with $J/\beta = 0.25$. In each panel, the dashed hexagon represents the supercell, and the solid or colored hexagons are the original moir\'e cells.}
    \label{fig:metastable}
\end{figure}

To further understand this frustration, we examine several metastable 
configurations for the triangular system. 
We define a local twist angle $\vartheta(\rr)$ and a local lattice mismatch $\varepsilon(\rr)$ in terms of the local registry parameter $\dd(\rr)$ such that
\begin{subequations}
\begin{gather}
    \nnabla\cdot\dd(\rr) = 2 \lp[(1+\varepsilon(\rr)) \cos\vartheta(\rr) - 1\rp], \\\addlinespace
    \nnabla\times\dd(\rr) = 2(1+\varepsilon(\rr)) \sin\vartheta(\rr).
\end{gather}
\end{subequations}
These equations reduce to Eq. (\ref{eq:registryrigid}) in the rigid lattice case, 
and to $\nnabla\cdot\dd = 2\varepsilon(\rr)$ and $\nnabla\times\dd = 2\vartheta(\rr)$ in the small $\varepsilon$ and $\vartheta$ limit. 
In Figs. \ref{fig:metastable} (a)-(d) the twirl configurations are visualized directly by plotting the local twist angle $\vartheta(\rr)$ for the four lowest energy solutions that we found. Pockets of positive and negative local twist angles correspond to left and right twirls, respectively. We make the following main observations from these results. First, the configurations shown in Figs. \ref{fig:metastable} (a), (b) and (d) are maximally staggered, \textit{i.e.}, each triangular plaquette
(such as the green triangle shown in (a)) contains at least one left twirl and one right twirl.  We refer to plaquettes with this property as ``good" plaquettes since they
minimize the energy of nearest-neighbor antiferromagnetic Ising models on a triangular lattice \cite{wannier1950antiferromagnetism}. 
We also note that the configuration illustrated in Fig. \ref{fig:metastable}(d) is more costly than that in (c), even though (c) contains two ``bad" plaquettes. 
This suggests that coupling between further neighbors and/or more than 2 twirls is significant. Second, the intensities of the twirls vary. This is a qualitatively non-trivial feature and can be understood from the following considerations. 
In the small angle limit, 
\begin{equation}
    \vartheta(\rr) \approx \frac{\nnabla\times\dd}{2} = \frac{\nnabla\times(\dd_0 + \ddelta)}{2} \approx \theta + \frac{\nnabla\times\ddelta}{2}.
    \label{eq:constraint}
\end{equation} 
Green's theorem demands that for the periodic function $\ddelta(\rr)$, $\int_{\rm supercell} d^2\rr \;  \nnabla \times \ddelta = 0$, 
and hence that the spatial average of the local twist angle $\vartheta(\rr)$ over the periodic supercell ($2\sqrt{3} \times 2\sqrt{3}$ supercell in our case) must be equal to the global twist angle $\theta$. 
For the aligned ($\theta = 0$) case, $\vartheta (\rr)$ must therefore 
spatially average to $0$ over the periodic supercell, 
even for configurations with unbalanced left and right twirls [See Fig.~\ref{fig:metastable} (b)].  It follows that the intensities of left and right twirls must be different
to satisfy the spatial average condition, and therefore that  
the classical Ising model is insufficient to fully capture the configuration energies.

We conclude that the discrete up/down pseudospins of the chiral model
should be replaced with continuous variables.  We propose that the behavior of twirls can be more accurately described by an antiferromagnetic lattice $\phi^4$ model with the energy functional 
\begin{equation}
    E = \sum_i \lp( \alpha\phi_i^4 - \beta\phi_i^2 \rp) + \sum_{i<j} J_{ij} \phi_i \phi_j,
    \label{eq:AFM_phi4}
\end{equation}
such that 
\begin{equation}
    \phi_i = \int_{{\rm MC}_i} \frac{\nnabla\times\dd(\rr)}{2} d^2\rr = \int_{{\rm MC}_i}\bigl( 1+\varepsilon(\rr) \bigr) \sin\vartheta(\rr) d^2\rr
    \label{eq:twirlmoment_def}
\end{equation}
is the \textit{twirl moment} in the $i$th moir\'e cell (${\rm MC}_i$). Note that the parameters $\alpha$ and $\beta$ can be eliminated by scaling $E$ and $\phi_i$, leaving only $J_{ij}/\beta$ as the parameters defining the $\phi^4$ model.
In Figs. \ref{fig:metastable}(e) and (f), we respectively show the twirl moment configuration of the lowest-energy state of the elasticity model and the effective $\phi^4$ model with nearest-neighbor coupling $J_{\Mean{ij}} = J = 0.25\beta$ under the global twist constraint $\sum_i \phi_i = 0$. Comparing the two, we see that the pattern of relative twirl intensities can be well reproduced by the $\phi^4$ model. 
Adding further-neighbor couplings can reproduce the patterns of all four metastable states shown in Figs. \ref{fig:metastable} (a)-(d) with the correct energy hierarchy \cite{supplemental}. 

\begin{figure}[t]
	\centering
	\includegraphics[width=0.48\textwidth]{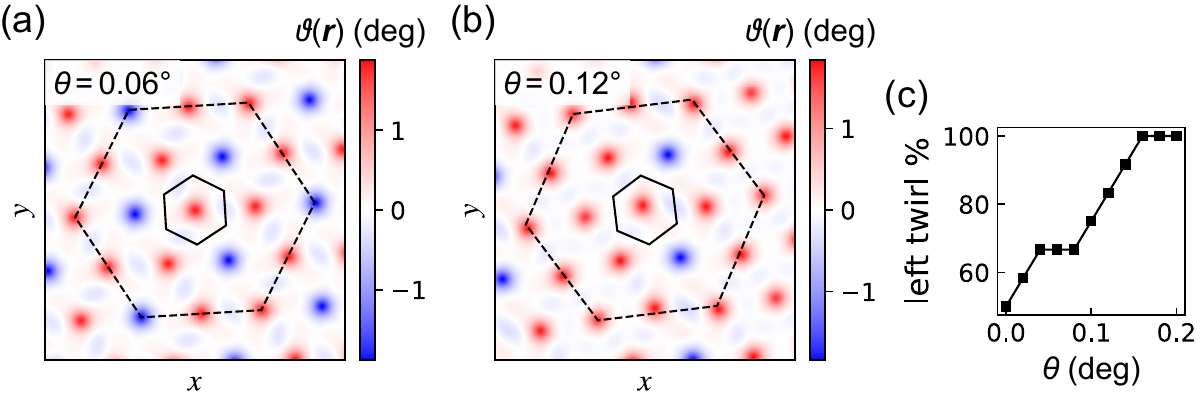}
	\caption{(a)-(b) The local twist angle plots of the lowest energy configuration of the triangular moir\'e system with the same parameter as plotted in Fig. \ref{fig:metastable}, but with finite global twist angles $\theta$ indicated on the top left. (c) The percentage of left (counterclockwise) twirls in all twirls, as a function of the global twist angle.}
	\label{fig:twist_evolution}
\end{figure}

Next, we examine the effect of the global twist on the twirl configurations. 
In Figs. \ref{fig:twist_evolution} (a) and (b), we present two representative examples of finite-$\theta$ configurations. 
In Fig. \ref{fig:twist_evolution} (c), we show that the fraction of the left twirls generally grows linearly with the global twist angle, with a pronounced plateau between $0.04^\circ \sim 0.08^\circ$ due to finite system size. This dependence is consistent with the requirement of the
aforementioned Green's theorem constraint that the average twirl moment equals the global twist angle (in the small angle limit and up to a factor of moir\'e cell area).
In contrast, for the square lattice, the left twirl fraction jumps directly from 50\% to 100\% at a finite twist angle (shown in the End Matter).
In both systems, the effective $\phi^4$ model captures these behaviors \cite{supplemental}.

\smallskip
\paragraph{Hysteresis with twist angle.}

As we just saw, twirls can be biased by applying a global twist, similar to spins being polarized by a magnetic field. In the presence of frustration, we can expect this reorientation process to be hysteretic \cite{maignan2000single, hardy2004temperature}.
To test this hypothesis,  we study the  
evolution of the structural twirls while slowly changing the global twist angle from $-0.2^\circ$ to $0.2^\circ$. 
We obtain the initial state (at $\theta = -0.2^\circ$) by iteration starting from a random initial $\ddelta(\rr)$, and obtain all subsequent states by iterating from the convergent result of the previous $\theta$, thus emulating an adiabatic evolution
\footnote{Note the subtlety that the size and orientation of moir\'e cell changes with the twist angle. Iterating from the exact same $\ddelta(\rr)$ at each $\rr$ point of the previous sample would suffer difficulties due to incompatibility in superlattice periodicity. Therefore, at each inter-sample transition, we scale our spatial grid to fit the next supercell without changing the magnitudes or the directions of $\ddelta(\rr)$'s on the grid points. This is expected to be a good approximation in the adiabatic limit.}.

\begin{figure}
    \centering
    \includegraphics[width=0.48\textwidth]{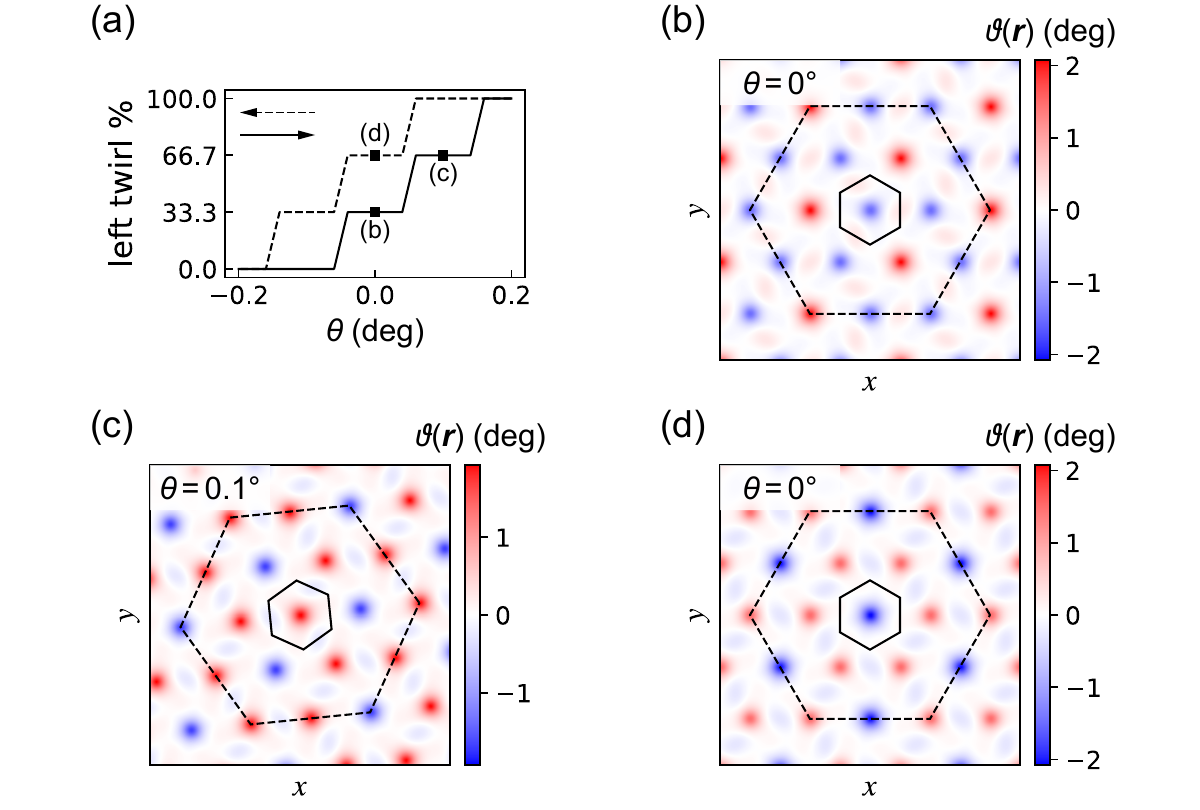}
    \caption{(a) Evolution of the left twirl percentage along a loop of $\theta$ between $-0.2^\circ$ and $0.2^\circ$ in increments of $\pm 0.02^\circ$. Forward and backward sweeps are respectively represented by solid and dashed lines. (b)-(d) The local twist angle plots of the system at the sample points marked in (a). The global twist angle of each sample is indicated on the top left.}
    \label{fig:hysteresis}
\end{figure}

As Fig. \ref{fig:hysteresis} (a) shows, we observe hysteretic behavior and stepwise evolution of the ratio of left twirls with plateaus at 1/3 and 2/3. The plateau states here display Kekul\'e-like honeycomb configurations of twirls with $\sqrt{3} \times \sqrt{3}$ supercells, as shown in Figs. \ref{fig:hysteresis} (b)-(d).
These behaviors strongly resemble the stepwise magnetization evolution and hysteresis under slowly varying external fields
in triangular lattice Ising antiferromagnets~\cite{kageyama1997fieldinduced, kudasov2006steplike, kudasov2008dynamics, yao2006monte}. The 3-step hysteresis between equidistant plateaus, as well as the Kekul\'e twirl configuration, is also observed in our effective $\phi^4$ model with an 
enlarged ($4\sqrt3 \times 4\sqrt{3}$) supercell \cite{supplemental}. 
We note that though challenging, such hysteresis studies may be feasible using 
recently demonstrated \textit{in situ} twistable moir\'e devices \cite{ribeiro-palau2018twistable, inbar2023quantum, farrar2025impact}. We also expect hysteretic response to variation in other control parameters that are more experimentally accessible, for example to gate voltages which will alter the stacking energy density function $V(\dd)$.

\smallskip
\paragraph{Summary and discussion.} 

In summary, using a simple continuum elasticity model for relaxed moir\'e heterobilayers, 
we have demonstrated the emergence of structural twirls in the domain wall junctions, with a dominant nearest neighbor antiferromagnetic coupling between the twirl directions. 
For the honeycomb lattice moir\'e, the domain wall network takes a triangular form, leading to structural frustration.  For the square lattice system, we obtain a simple staggered antiferromagnetic ground state.
To accurately capture the low-energy metastable states in both cases we constructed an effective $\phi^4$ model. 
The model also allows for simulations of systems of a much larger size \cite{supplemental}. 

We now comment on realistic model parameters for the twirls and frustration phenomena to occur in real moir\'e systems.
As we have argued, twirling is expected in the nonlinear regime of Eq. (\ref{eq:relax}), which occurs for strong enough interaction $V_1$ that is at least the same order as $\mu\epsilon^2$. For the model parameters we use in our calculations, $\epsilon = -0.017$ and $V_1/\mu = 3\times 10^{-4}$, meaning that $V_1/\mu\epsilon^2 \approx 1$ is sufficient for twirling.
Heterobilayer TMDs with the same type of chalcogen atom have extremely small lattice mismatches ($\epsilon\sim10^{-3}$) \cite{mesple2023giant, kaliteevski2023twirling}, conveniently satisfying the above condition. Therefore, the heterobilayer TMDs are likely to support frustrated twirls in aligned devices. 
In realistic systems, however, there will inevitably be supermoir\'e-scale variations of twist angles. Due to the small threshold twist angle at which all twirls are aligned, some of the patches may show aligned twirls, while others may show Kekul\'e-like $\sqrt{3} \times \sqrt{3}$ twirl configurations. Frustrated patterns may only be observable within transitional regions between these patches, where layers happen to be near-perfectly aligned.
On the other hand, TMD heterobilayers with
different types of chalcogen atoms generally have $|\epsilon|\sim0.04$ \cite{haastrup2018computational, gjerding2021recent}, suggesting lower sensitivity to twist angle variations, but also requiring higher thresholds in $V_1/\mu$. The threshold can potentially be lowered and/or reached by engineering with heterostrain, pressure and/or electric field.

As the lattice relaxation modifies the background potential experienced by electrons, it will influence the electronic properties in moir\'e materials \cite{nam2017lattice, soltero2024competition}. 
For instance, the ``frustrated and frozen'' distorted domain structures can act as a disordered potential for electrons, favoring Anderson localization \cite{anderson1958absence}. 
Further, the coupling between the twirl distortions and electrons provides a novel emergent mechanism of electron-lattice coupling in moir\'e materials, which could play a role in the widely observed hysteretic behaviors in electronic transport properties in response to gate-tuned chemical potential and electric fields 
\cite{yasuda2021stackingengineered, zheng2020unconventional, niu2022giant, zheng2023electronic, zhang2024electronic, yan2023moire, waters2025anomalous, niu2025ferroelectricity}. 
We leave the study of these effects for the future.

\section{End Matter}
\begin{figure}
    \centering
    \includegraphics[width=0.48\textwidth]{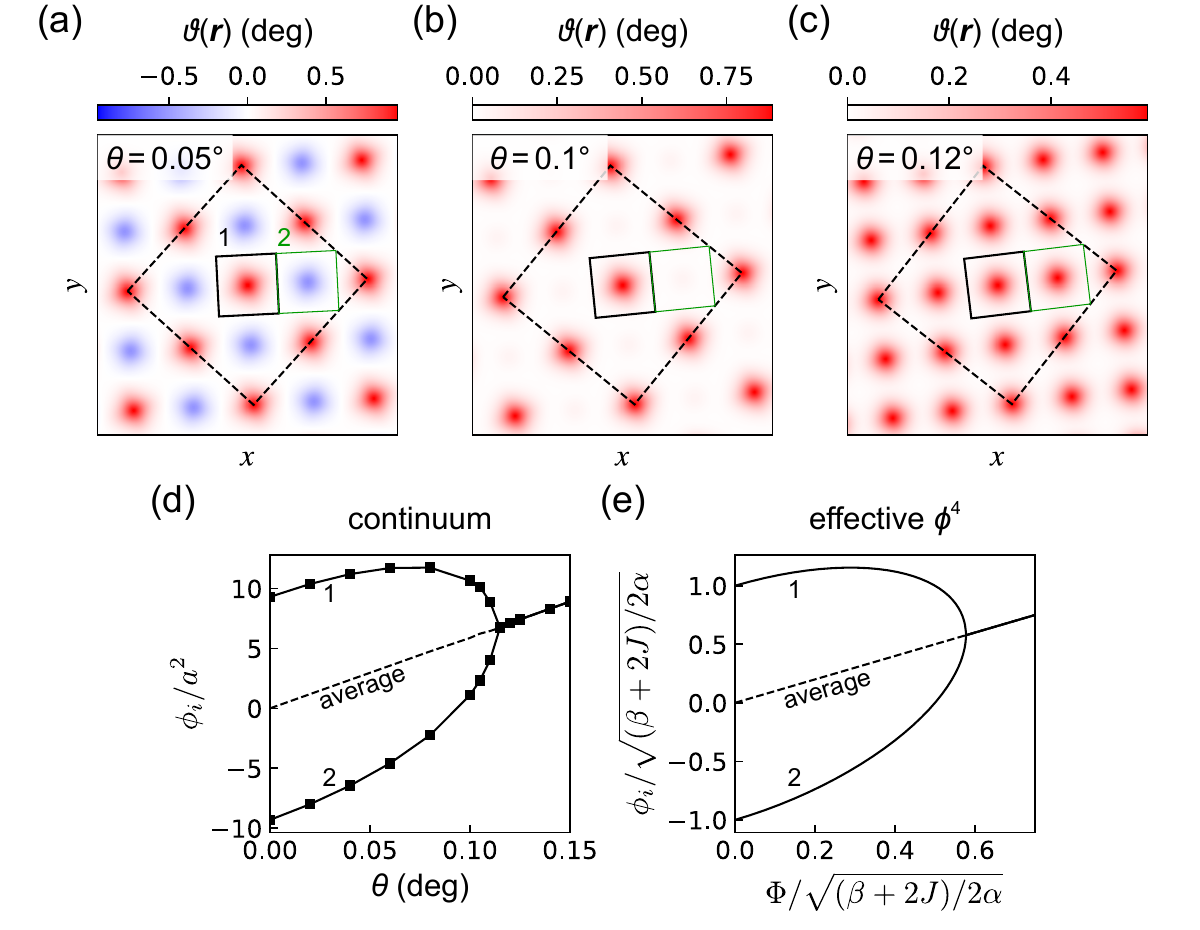}
    \caption{(a)-(c) The local twist angle plots of the lowest energy configuration of a square system with the same parameter as plotted in Fig. \ref{fig:introduction} (b) except that a finite global twist angle $\theta$ is introduced, as indicated on the top left of each panel. (d) Twist angle dependence of the twirl moments in the two moir\'e cells marked with solid black (1) and green (2) squares in (a)-(c). (e) Dependence of the individual twirl moments $\phi_i$ on the average twirl moment $\Phi$, in the analytical formulation of the effective $\phi^4$ model, where only nearest-neighbor coupling $J = J_{\Mean{ij}}$ is included..}
    \label{figE:square1}
\end{figure}

The behaviors of the twirls in the square system are drastically different than in the triangular system, yet we show that our effective $\phi^4$ model well describes the square system. As Figs. \ref{figE:square1} (a)-(c) show, in a square system with $\epsilon = -0.017$ and $V_1/\mu = 3\times 10^{-4}$, the right/clockwise twirls (corresponding to the blue spots) continuously shrink as the global twist angle increases, and reverse at around $\theta = 0.1^\circ$. At a critical angle around $0.115^\circ$, all twirls align with equal intensity, restoring the original square-lattice moir\'e periodicity. The twirl moments in the square moir\'e cells marked in Fig. \ref{figE:square1} (a)-(c) are extracted in the same way as Eq. (\ref{eq:twirlmoment_def}), and shown in Fig. \ref{figE:square1} (d). Moreover, the N\'eel states in the antiferromagnetic square-lattice $\phi^4$ model can be analytically solved \cite{supplemental}. Fig. \ref{figE:square1} (e) shows the analytical solution of the $\phi^4$ model with pure nearest-neighbor coupling, which accurately reproduces the shape of Fig. \ref{figE:square1} (d). This further supports the validity of our effective $\phi^4$ model.

\begin{figure}
    \centering
    \includegraphics[width=0.48\textwidth]{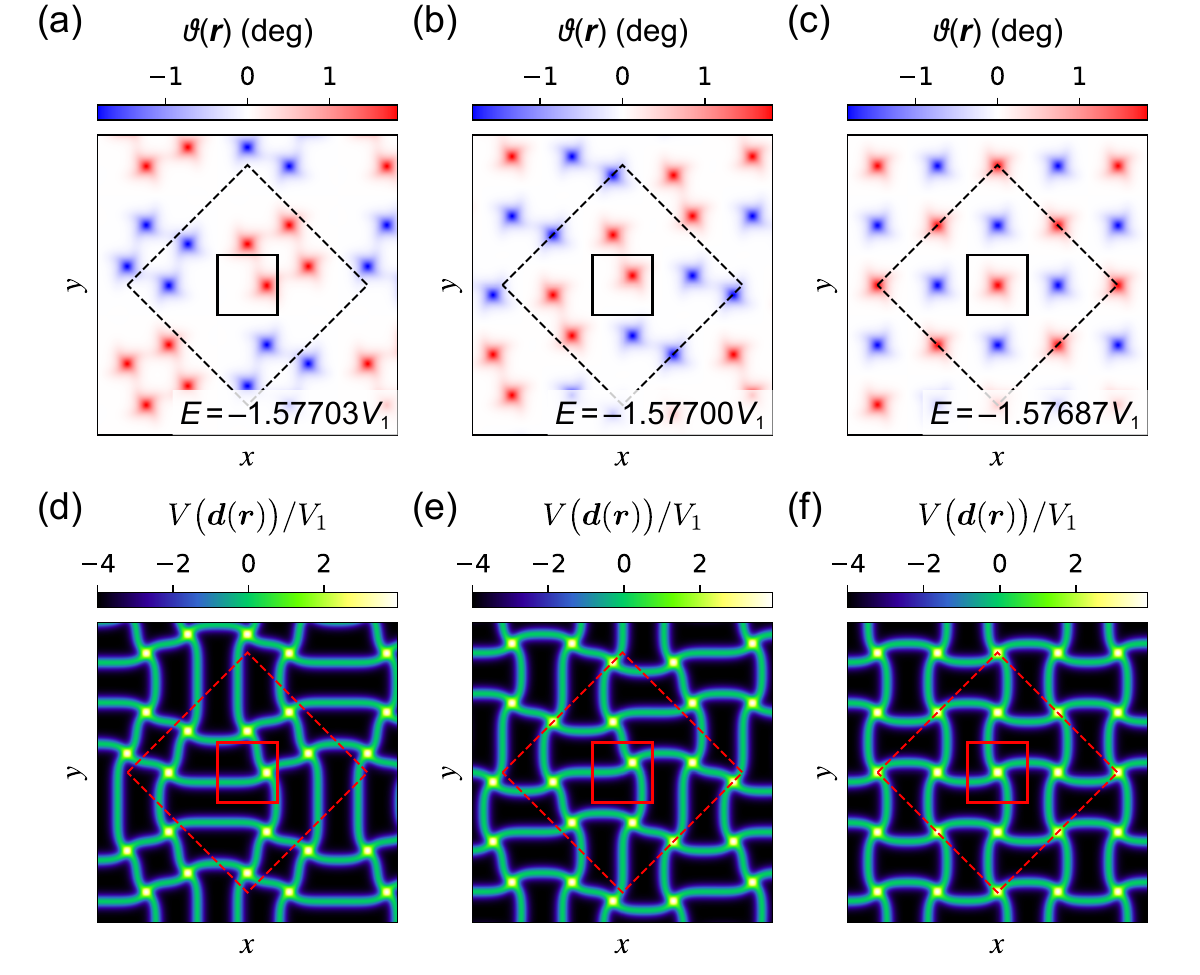}
    \caption{(a)-(c) Local twist angle plots of several metastable states in a square system with lattice mismatch $\epsilon = -0.01$, $\theta = 0$, $\lambda/\mu = 0.5$ and $V_1/\mu = 3\times 10^{-4}$. The total energy per area is marked at the bottom right of each plot. (d)-(f) The corresponding local stacking energy plots of these metastable states.}
    \label{figE:square2}
\end{figure}

We also show a case that highlights the limitation of the $\phi^4$ model.
Fig. \ref{figE:square2} shows an example of square lattice results with $\epsilon = -0.01$. We see two types of extra metastable states, indicated in Figs. \ref{figE:square2} (a)-(b), which are more stable than the staggered configuration shown in Fig. \ref{figE:square2} (c). The former two feature chirality-aligned tetramers and dimers of twirls, respectively, not to exclude possible more complex patterns if allowing for further translational symmetry breaking than our $2\sqrt{2} \times 2\sqrt{2}$ supercells. 
The existence of these states suggests that further refinements in the effective lattice $\phi^4$ model should involve including the displacement of the lattice sites as another degree of freedom and introducing displacement-dependent coupling functions. 
Similar effects in triangular systems featuring trimers of twirls have been reported in atomistic simulations \cite{maity2021reconstruction}. 
However, we do not observe such trimer features in triangular systems.

{\it Acknowledgements:}
We acknowledge computational resources provided by the Texas Advanced Computing Center (TACC).
J.S. and I.M. acknowledge support by the US Department of Energy, Office of Science, Basic Energy Sciences, Materials Sciences and Engineering Division. 
G.C. acknowledges support from Engineering and Physical Sciences Research Council (EPSRC) grant EP/V062654/1 and support by grant NSF PHY-2309135 to the Kavli Institute for Theoretical Physics during the initial part of this work. 
We are thankful to Alejandro Ramos-Alonso, Daniel Bennett and Alex Watson for helpful interactions.

J.S. and I.M. conceptualized the project. J.S. performed the technical calculations with guidance from I.M. All authors contributed to discussion and writing of the manuscript

\bibliography{bibliography}

\begin{thebibliography}{52}%
\makeatletter
\providecommand \@ifxundefined [1]{%
 \@ifx{#1\undefined}
}%
\providecommand \@ifnum [1]{%
 \ifnum #1\expandafter \@firstoftwo
 \else \expandafter \@secondoftwo
 \fi
}%
\providecommand \@ifx [1]{%
 \ifx #1\expandafter \@firstoftwo
 \else \expandafter \@secondoftwo
 \fi
}%
\providecommand \natexlab [1]{#1}%
\providecommand \enquote  [1]{``#1''}%
\providecommand \bibnamefont  [1]{#1}%
\providecommand \bibfnamefont [1]{#1}%
\providecommand \citenamefont [1]{#1}%
\providecommand \href@noop [0]{\@secondoftwo}%
\providecommand \href [0]{\begingroup \@sanitize@url \@href}%
\providecommand \@href[1]{\@@startlink{#1}\@@href}%
\providecommand \@@href[1]{\endgroup#1\@@endlink}%
\providecommand \@sanitize@url [0]{\catcode `\\12\catcode `\$12\catcode
  `\&12\catcode `\#12\catcode `\^12\catcode `\_12\catcode `\%12\relax}%
\providecommand \@@startlink[1]{}%
\providecommand \@@endlink[0]{}%
\providecommand \url  [0]{\begingroup\@sanitize@url \@url }%
\providecommand \@url [1]{\endgroup\@href {#1}{\urlprefix }}%
\providecommand \urlprefix  [0]{URL }%
\providecommand \Eprint [0]{\href }%
\providecommand \doibase [0]{https://doi.org/}%
\providecommand \selectlanguage [0]{\@gobble}%
\providecommand \bibinfo  [0]{\@secondoftwo}%
\providecommand \bibfield  [0]{\@secondoftwo}%
\providecommand \translation [1]{[#1]}%
\providecommand \BibitemOpen [0]{}%
\providecommand \bibitemStop [0]{}%
\providecommand \bibitemNoStop [0]{.\EOS\space}%
\providecommand \EOS [0]{\spacefactor3000\relax}%
\providecommand \BibitemShut  [1]{\csname bibitem#1\endcsname}%
\let\auto@bib@innerbib\@empty
\bibitem [{\citenamefont {Cao}\ \emph {et~al.}(2018{\natexlab{a}})\citenamefont
  {Cao}, \citenamefont {Fatemi}, \citenamefont {Demir}, \citenamefont {Fang},
  \citenamefont {Tomarken}, \citenamefont {Luo}, \citenamefont
  {{Sanchez-Yamagishi}}, \citenamefont {Watanabe}, \citenamefont {Taniguchi},
  \citenamefont {Kaxiras}, \citenamefont {Ashoori},\ and\ \citenamefont
  {{Jarillo-Herrero}}}]{cao2018correlated}%
  \BibitemOpen
  \bibfield  {author} {\bibinfo {author} {\bibfnamefont {Y.}~\bibnamefont
  {Cao}}, \bibinfo {author} {\bibfnamefont {V.}~\bibnamefont {Fatemi}},
  \bibinfo {author} {\bibfnamefont {A.}~\bibnamefont {Demir}}, \bibinfo
  {author} {\bibfnamefont {S.}~\bibnamefont {Fang}}, \bibinfo {author}
  {\bibfnamefont {S.~L.}\ \bibnamefont {Tomarken}}, \bibinfo {author}
  {\bibfnamefont {J.~Y.}\ \bibnamefont {Luo}}, \bibinfo {author} {\bibfnamefont
  {J.~D.}\ \bibnamefont {{Sanchez-Yamagishi}}}, \bibinfo {author}
  {\bibfnamefont {K.}~\bibnamefont {Watanabe}}, \bibinfo {author}
  {\bibfnamefont {T.}~\bibnamefont {Taniguchi}}, \bibinfo {author}
  {\bibfnamefont {E.}~\bibnamefont {Kaxiras}}, \bibinfo {author} {\bibfnamefont
  {R.~C.}\ \bibnamefont {Ashoori}},\ and\ \bibinfo {author} {\bibfnamefont
  {P.}~\bibnamefont {{Jarillo-Herrero}}},\ }\bibfield  {title} {\bibinfo
  {title} {Correlated insulator behaviour at half-filling in magic-angle
  graphene superlattices},\ }\href {https://doi.org/10.1038/nature26154}
  {\bibfield  {journal} {\bibinfo  {journal} {Nature}\ }\textbf {\bibinfo
  {volume} {556}},\ \bibinfo {pages} {80} (\bibinfo {year}
  {2018}{\natexlab{a}})}\BibitemShut {NoStop}%
\bibitem [{\citenamefont {Cao}\ \emph {et~al.}(2018{\natexlab{b}})\citenamefont
  {Cao}, \citenamefont {Fatemi}, \citenamefont {Fang}, \citenamefont
  {Watanabe}, \citenamefont {Taniguchi}, \citenamefont {Kaxiras},\ and\
  \citenamefont {{Jarillo-Herrero}}}]{cao2018unconventional}%
  \BibitemOpen
  \bibfield  {author} {\bibinfo {author} {\bibfnamefont {Y.}~\bibnamefont
  {Cao}}, \bibinfo {author} {\bibfnamefont {V.}~\bibnamefont {Fatemi}},
  \bibinfo {author} {\bibfnamefont {S.}~\bibnamefont {Fang}}, \bibinfo {author}
  {\bibfnamefont {K.}~\bibnamefont {Watanabe}}, \bibinfo {author}
  {\bibfnamefont {T.}~\bibnamefont {Taniguchi}}, \bibinfo {author}
  {\bibfnamefont {E.}~\bibnamefont {Kaxiras}},\ and\ \bibinfo {author}
  {\bibfnamefont {P.}~\bibnamefont {{Jarillo-Herrero}}},\ }\bibfield  {title}
  {\bibinfo {title} {Unconventional superconductivity in magic-angle graphene
  superlattices},\ }\href {https://doi.org/10.1038/nature26160} {\bibfield
  {journal} {\bibinfo  {journal} {Nature}\ }\textbf {\bibinfo {volume} {556}},\
  \bibinfo {pages} {43} (\bibinfo {year} {2018}{\natexlab{b}})}\BibitemShut
  {NoStop}%
\bibitem [{\citenamefont {Cai}\ \emph {et~al.}(2023)\citenamefont {Cai},
  \citenamefont {Anderson}, \citenamefont {Wang}, \citenamefont {Zhang},
  \citenamefont {Liu}, \citenamefont {Holtzmann}, \citenamefont {Zhang},
  \citenamefont {Fan}, \citenamefont {Taniguchi}, \citenamefont {Watanabe},
  \citenamefont {Ran}, \citenamefont {Cao}, \citenamefont {Fu}, \citenamefont
  {Xiao}, \citenamefont {Yao},\ and\ \citenamefont {Xu}}]{cai2023signatures}%
  \BibitemOpen
  \bibfield  {author} {\bibinfo {author} {\bibfnamefont {J.}~\bibnamefont
  {Cai}}, \bibinfo {author} {\bibfnamefont {E.}~\bibnamefont {Anderson}},
  \bibinfo {author} {\bibfnamefont {C.}~\bibnamefont {Wang}}, \bibinfo {author}
  {\bibfnamefont {X.}~\bibnamefont {Zhang}}, \bibinfo {author} {\bibfnamefont
  {X.}~\bibnamefont {Liu}}, \bibinfo {author} {\bibfnamefont {W.}~\bibnamefont
  {Holtzmann}}, \bibinfo {author} {\bibfnamefont {Y.}~\bibnamefont {Zhang}},
  \bibinfo {author} {\bibfnamefont {F.}~\bibnamefont {Fan}}, \bibinfo {author}
  {\bibfnamefont {T.}~\bibnamefont {Taniguchi}}, \bibinfo {author}
  {\bibfnamefont {K.}~\bibnamefont {Watanabe}}, \bibinfo {author}
  {\bibfnamefont {Y.}~\bibnamefont {Ran}}, \bibinfo {author} {\bibfnamefont
  {T.}~\bibnamefont {Cao}}, \bibinfo {author} {\bibfnamefont {L.}~\bibnamefont
  {Fu}}, \bibinfo {author} {\bibfnamefont {D.}~\bibnamefont {Xiao}}, \bibinfo
  {author} {\bibfnamefont {W.}~\bibnamefont {Yao}},\ and\ \bibinfo {author}
  {\bibfnamefont {X.}~\bibnamefont {Xu}},\ }\bibfield  {title} {\bibinfo
  {title} {Signatures of fractional quantum anomalous {{Hall}} states in
  twisted {{MoTe2}}},\ }\href {https://doi.org/10.1038/s41586-023-06289-w}
  {\bibfield  {journal} {\bibinfo  {journal} {Nature}\ }\textbf {\bibinfo
  {volume} {622}},\ \bibinfo {pages} {63} (\bibinfo {year} {2023})}\BibitemShut
  {NoStop}%
\bibitem [{\citenamefont {Andrei}\ \emph {et~al.}(2021)\citenamefont {Andrei},
  \citenamefont {Efetov}, \citenamefont {{Jarillo-Herrero}}, \citenamefont
  {MacDonald}, \citenamefont {Mak}, \citenamefont {Senthil}, \citenamefont
  {Tutuc}, \citenamefont {Yazdani},\ and\ \citenamefont
  {Young}}]{andrei2021marvels}%
  \BibitemOpen
  \bibfield  {author} {\bibinfo {author} {\bibfnamefont {E.~Y.}\ \bibnamefont
  {Andrei}}, \bibinfo {author} {\bibfnamefont {D.~K.}\ \bibnamefont {Efetov}},
  \bibinfo {author} {\bibfnamefont {P.}~\bibnamefont {{Jarillo-Herrero}}},
  \bibinfo {author} {\bibfnamefont {A.~H.}\ \bibnamefont {MacDonald}}, \bibinfo
  {author} {\bibfnamefont {K.~F.}\ \bibnamefont {Mak}}, \bibinfo {author}
  {\bibfnamefont {T.}~\bibnamefont {Senthil}}, \bibinfo {author} {\bibfnamefont
  {E.}~\bibnamefont {Tutuc}}, \bibinfo {author} {\bibfnamefont
  {A.}~\bibnamefont {Yazdani}},\ and\ \bibinfo {author} {\bibfnamefont {A.~F.}\
  \bibnamefont {Young}},\ }\bibfield  {title} {\bibinfo {title} {The marvels of
  moir{\'e} materials},\ }\href {https://doi.org/10.1038/s41578-021-00284-1}
  {\bibfield  {journal} {\bibinfo  {journal} {Nat Rev Mater}\ }\textbf
  {\bibinfo {volume} {6}},\ \bibinfo {pages} {201} (\bibinfo {year}
  {2021})}\BibitemShut {NoStop}%
\bibitem [{\citenamefont {Carr}\ \emph {et~al.}(2018)\citenamefont {Carr},
  \citenamefont {Massatt}, \citenamefont {Torrisi}, \citenamefont {Cazeaux},
  \citenamefont {Luskin},\ and\ \citenamefont {Kaxiras}}]{carr2018relaxation}%
  \BibitemOpen
  \bibfield  {author} {\bibinfo {author} {\bibfnamefont {S.}~\bibnamefont
  {Carr}}, \bibinfo {author} {\bibfnamefont {D.}~\bibnamefont {Massatt}},
  \bibinfo {author} {\bibfnamefont {S.~B.}\ \bibnamefont {Torrisi}}, \bibinfo
  {author} {\bibfnamefont {P.}~\bibnamefont {Cazeaux}}, \bibinfo {author}
  {\bibfnamefont {M.}~\bibnamefont {Luskin}},\ and\ \bibinfo {author}
  {\bibfnamefont {E.}~\bibnamefont {Kaxiras}},\ }\bibfield  {title} {\bibinfo
  {title} {Relaxation and domain formation in incommensurate two-dimensional
  heterostructures},\ }\href {https://doi.org/10.1103/PhysRevB.98.224102}
  {\bibfield  {journal} {\bibinfo  {journal} {Phys. Rev. B}\ }\textbf {\bibinfo
  {volume} {98}},\ \bibinfo {pages} {224102} (\bibinfo {year}
  {2018})}\BibitemShut {NoStop}%
\bibitem [{\citenamefont {Alden}\ \emph {et~al.}(2013)\citenamefont {Alden},
  \citenamefont {Tsen}, \citenamefont {Huang}, \citenamefont {Hovden},
  \citenamefont {Brown}, \citenamefont {Park}, \citenamefont {Muller},\ and\
  \citenamefont {McEuen}}]{alden2013strain}%
  \BibitemOpen
  \bibfield  {author} {\bibinfo {author} {\bibfnamefont {J.~S.}\ \bibnamefont
  {Alden}}, \bibinfo {author} {\bibfnamefont {A.~W.}\ \bibnamefont {Tsen}},
  \bibinfo {author} {\bibfnamefont {P.~Y.}\ \bibnamefont {Huang}}, \bibinfo
  {author} {\bibfnamefont {R.}~\bibnamefont {Hovden}}, \bibinfo {author}
  {\bibfnamefont {L.}~\bibnamefont {Brown}}, \bibinfo {author} {\bibfnamefont
  {J.}~\bibnamefont {Park}}, \bibinfo {author} {\bibfnamefont {D.~A.}\
  \bibnamefont {Muller}},\ and\ \bibinfo {author} {\bibfnamefont {P.~L.}\
  \bibnamefont {McEuen}},\ }\bibfield  {title} {\bibinfo {title} {Strain
  solitons and topological defects in bilayer graphene},\ }\href
  {https://doi.org/10.1073/pnas.1309394110} {\bibfield  {journal} {\bibinfo
  {journal} {Proceedings of the National Academy of Sciences}\ }\textbf
  {\bibinfo {volume} {110}},\ \bibinfo {pages} {11256} (\bibinfo {year}
  {2013})}\BibitemShut {NoStop}%
\bibitem [{\citenamefont {Yoo}\ \emph {et~al.}(2019)\citenamefont {Yoo},
  \citenamefont {Engelke}, \citenamefont {Carr}, \citenamefont {Fang},
  \citenamefont {Zhang}, \citenamefont {Cazeaux}, \citenamefont {Sung},
  \citenamefont {Hovden}, \citenamefont {Tsen}, \citenamefont {Taniguchi} \emph
  {et~al.}}]{yoo2019atomic}%
  \BibitemOpen
  \bibfield  {author} {\bibinfo {author} {\bibfnamefont {H.}~\bibnamefont
  {Yoo}}, \bibinfo {author} {\bibfnamefont {R.}~\bibnamefont {Engelke}},
  \bibinfo {author} {\bibfnamefont {S.}~\bibnamefont {Carr}}, \bibinfo {author}
  {\bibfnamefont {S.}~\bibnamefont {Fang}}, \bibinfo {author} {\bibfnamefont
  {K.}~\bibnamefont {Zhang}}, \bibinfo {author} {\bibfnamefont
  {P.}~\bibnamefont {Cazeaux}}, \bibinfo {author} {\bibfnamefont {S.~H.}\
  \bibnamefont {Sung}}, \bibinfo {author} {\bibfnamefont {R.}~\bibnamefont
  {Hovden}}, \bibinfo {author} {\bibfnamefont {A.~W.}\ \bibnamefont {Tsen}},
  \bibinfo {author} {\bibfnamefont {T.}~\bibnamefont {Taniguchi}}, \emph
  {et~al.},\ }\bibfield  {title} {\bibinfo {title} {Atomic and electronic
  reconstruction at the van der {{Waals}} interface in twisted bilayer
  graphene},\ }\href {https://doi.org/10.1038/s41563-019-0346-z} {\bibfield
  {journal} {\bibinfo  {journal} {Nat. Mater.}\ }\textbf {\bibinfo {volume}
  {18}},\ \bibinfo {pages} {448} (\bibinfo {year} {2019})}\BibitemShut
  {NoStop}%
\bibitem [{\citenamefont {Weston}\ \emph {et~al.}(2020)\citenamefont {Weston},
  \citenamefont {Zou}, \citenamefont {Enaldiev}, \citenamefont {Summerfield},
  \citenamefont {Clark}, \citenamefont {Z{\'o}lyomi}, \citenamefont {Graham},
  \citenamefont {Yelgel}, \citenamefont {Magorrian}, \citenamefont {Zhou} \emph
  {et~al.}}]{weston2020atomic}%
  \BibitemOpen
  \bibfield  {author} {\bibinfo {author} {\bibfnamefont {A.}~\bibnamefont
  {Weston}}, \bibinfo {author} {\bibfnamefont {Y.}~\bibnamefont {Zou}},
  \bibinfo {author} {\bibfnamefont {V.}~\bibnamefont {Enaldiev}}, \bibinfo
  {author} {\bibfnamefont {A.}~\bibnamefont {Summerfield}}, \bibinfo {author}
  {\bibfnamefont {N.}~\bibnamefont {Clark}}, \bibinfo {author} {\bibfnamefont
  {V.}~\bibnamefont {Z{\'o}lyomi}}, \bibinfo {author} {\bibfnamefont
  {A.}~\bibnamefont {Graham}}, \bibinfo {author} {\bibfnamefont
  {C.}~\bibnamefont {Yelgel}}, \bibinfo {author} {\bibfnamefont
  {S.}~\bibnamefont {Magorrian}}, \bibinfo {author} {\bibfnamefont
  {M.}~\bibnamefont {Zhou}}, \emph {et~al.},\ }\bibfield  {title} {\bibinfo
  {title} {Atomic reconstruction in twisted bilayers of transition metal
  dichalcogenides},\ }\href {https://doi.org/10.1038/s41565-020-0682-9}
  {\bibfield  {journal} {\bibinfo  {journal} {Nat. Nanotechnol.}\ }\textbf
  {\bibinfo {volume} {15}},\ \bibinfo {pages} {592} (\bibinfo {year}
  {2020})}\BibitemShut {NoStop}%
\bibitem [{\citenamefont {Molino}\ \emph {et~al.}(2023)\citenamefont {Molino},
  \citenamefont {Aggarwal}, \citenamefont {Enaldiev}, \citenamefont
  {Plumadore}, \citenamefont {I.~Fal{\textasciiacute}ko},\ and\ \citenamefont
  {Luican-Mayer}}]{molino2023ferroelectric}%
  \BibitemOpen
  \bibfield  {author} {\bibinfo {author} {\bibfnamefont {L.}~\bibnamefont
  {Molino}}, \bibinfo {author} {\bibfnamefont {L.}~\bibnamefont {Aggarwal}},
  \bibinfo {author} {\bibfnamefont {V.}~\bibnamefont {Enaldiev}}, \bibinfo
  {author} {\bibfnamefont {R.}~\bibnamefont {Plumadore}}, \bibinfo {author}
  {\bibfnamefont {V.}~\bibnamefont {I.~Fal{\textasciiacute}ko}},\ and\ \bibinfo
  {author} {\bibfnamefont {A.}~\bibnamefont {Luican-Mayer}},\ }\bibfield
  {title} {\bibinfo {title} {Ferroelectric {{Switching}} at
  {{Symmetry}}-{{Broken Interfaces}} by {{Local Control}} of {{Dislocations
  Networks}}},\ }\href {https://doi.org/10.1002/adma.202207816} {\bibfield
  {journal} {\bibinfo  {journal} {Advanced Materials}\ }\textbf {\bibinfo
  {volume} {35}},\ \bibinfo {pages} {2207816} (\bibinfo {year}
  {2023})}\BibitemShut {NoStop}%
\bibitem [{\citenamefont {McGilly}\ \emph {et~al.}(2020)\citenamefont
  {McGilly}, \citenamefont {Kerelsky}, \citenamefont {Finney}, \citenamefont
  {Shapovalov}, \citenamefont {Shih}, \citenamefont {Ghiotto}, \citenamefont
  {Zeng}, \citenamefont {Moore}, \citenamefont {Wu}, \citenamefont {Bai} \emph
  {et~al.}}]{Mcgilly2020visualization}%
  \BibitemOpen
  \bibfield  {author} {\bibinfo {author} {\bibfnamefont {L.~J.}\ \bibnamefont
  {McGilly}}, \bibinfo {author} {\bibfnamefont {A.}~\bibnamefont {Kerelsky}},
  \bibinfo {author} {\bibfnamefont {N.~R.}\ \bibnamefont {Finney}}, \bibinfo
  {author} {\bibfnamefont {K.}~\bibnamefont {Shapovalov}}, \bibinfo {author}
  {\bibfnamefont {E.-M.}\ \bibnamefont {Shih}}, \bibinfo {author}
  {\bibfnamefont {A.}~\bibnamefont {Ghiotto}}, \bibinfo {author} {\bibfnamefont
  {Y.}~\bibnamefont {Zeng}}, \bibinfo {author} {\bibfnamefont {S.~L.}\
  \bibnamefont {Moore}}, \bibinfo {author} {\bibfnamefont {W.}~\bibnamefont
  {Wu}}, \bibinfo {author} {\bibfnamefont {Y.}~\bibnamefont {Bai}}, \emph
  {et~al.},\ }\bibfield  {title} {\bibinfo {title} {Visualization of moir{\'e}
  superlattices},\ }\href {https://doi.org/10.1038/s41565-020-0708-3}
  {\bibfield  {journal} {\bibinfo  {journal} {Nat. Nanotechnol.}\ }\textbf
  {\bibinfo {volume} {15}},\ \bibinfo {pages} {580} (\bibinfo {year}
  {2020})}\BibitemShut {NoStop}%
\bibitem [{\citenamefont {Dai}\ \emph {et~al.}(2016)\citenamefont {Dai},
  \citenamefont {Xiang},\ and\ \citenamefont {Srolovitz}}]{dai2016twisted}%
  \BibitemOpen
  \bibfield  {author} {\bibinfo {author} {\bibfnamefont {S.}~\bibnamefont
  {Dai}}, \bibinfo {author} {\bibfnamefont {Y.}~\bibnamefont {Xiang}},\ and\
  \bibinfo {author} {\bibfnamefont {D.~J.}\ \bibnamefont {Srolovitz}},\
  }\bibfield  {title} {\bibinfo {title} {Twisted {{Bilayer Graphene}}:
  {{Moir{\'e}}} with a {{Twist}}},\ }\href
  {https://doi.org/10.1021/acs.nanolett.6b02870} {\bibfield  {journal}
  {\bibinfo  {journal} {Nano Lett.}\ }\textbf {\bibinfo {volume} {16}},\
  \bibinfo {pages} {5923} (\bibinfo {year} {2016})}\BibitemShut {NoStop}%
\bibitem [{\citenamefont {Quan}\ \emph {et~al.}(2018)\citenamefont {Quan},
  \citenamefont {He},\ and\ \citenamefont {Ni}}]{quan2018tunable}%
  \BibitemOpen
  \bibfield  {author} {\bibinfo {author} {\bibfnamefont {S.}~\bibnamefont
  {Quan}}, \bibinfo {author} {\bibfnamefont {L.}~\bibnamefont {He}},\ and\
  \bibinfo {author} {\bibfnamefont {Y.}~\bibnamefont {Ni}},\ }\bibfield
  {title} {\bibinfo {title} {Tunable mosaic structures in van der {{Waals}}
  layered materials},\ }\href {https://doi.org/10.1039/C8CP04360D} {\bibfield
  {journal} {\bibinfo  {journal} {Phys. Chem. Chem. Phys.}\ }\textbf {\bibinfo
  {volume} {20}},\ \bibinfo {pages} {25428} (\bibinfo {year}
  {2018})}\BibitemShut {NoStop}%
\bibitem [{\citenamefont {Maity}\ \emph {et~al.}(2021)\citenamefont {Maity},
  \citenamefont {Maiti}, \citenamefont {Krishnamurthy},\ and\ \citenamefont
  {Jain}}]{maity2021reconstruction}%
  \BibitemOpen
  \bibfield  {author} {\bibinfo {author} {\bibfnamefont {I.}~\bibnamefont
  {Maity}}, \bibinfo {author} {\bibfnamefont {P.~K.}\ \bibnamefont {Maiti}},
  \bibinfo {author} {\bibfnamefont {H.~R.}\ \bibnamefont {Krishnamurthy}},\
  and\ \bibinfo {author} {\bibfnamefont {M.}~\bibnamefont {Jain}},\ }\bibfield
  {title} {\bibinfo {title} {Reconstruction of moir{\'e} lattices in twisted
  transition metal dichalcogenide bilayers},\ }\href
  {https://doi.org/10.1103/PhysRevB.103.L121102} {\bibfield  {journal}
  {\bibinfo  {journal} {Phys. Rev. B}\ }\textbf {\bibinfo {volume} {103}},\
  \bibinfo {pages} {L121102} (\bibinfo {year} {2021})}\BibitemShut {NoStop}%
\bibitem [{\citenamefont {Mesple}\ \emph {et~al.}(2023)\citenamefont {Mesple},
  \citenamefont {Walet}, \citenamefont {Trambly De~Laissardi{\`e}re},
  \citenamefont {Guinea}, \citenamefont {Do{\v s}enovi{\'c}}, \citenamefont
  {Okuno}, \citenamefont {Paillet}, \citenamefont {Michon}, \citenamefont
  {Chapelier},\ and\ \citenamefont {Renard}}]{mesple2023giant}%
  \BibitemOpen
  \bibfield  {author} {\bibinfo {author} {\bibfnamefont {F.}~\bibnamefont
  {Mesple}}, \bibinfo {author} {\bibfnamefont {N.~R.}\ \bibnamefont {Walet}},
  \bibinfo {author} {\bibfnamefont {G.}~\bibnamefont {Trambly
  De~Laissardi{\`e}re}}, \bibinfo {author} {\bibfnamefont {F.}~\bibnamefont
  {Guinea}}, \bibinfo {author} {\bibfnamefont {D.}~\bibnamefont {Do{\v
  s}enovi{\'c}}}, \bibinfo {author} {\bibfnamefont {H.}~\bibnamefont {Okuno}},
  \bibinfo {author} {\bibfnamefont {C.}~\bibnamefont {Paillet}}, \bibinfo
  {author} {\bibfnamefont {A.}~\bibnamefont {Michon}}, \bibinfo {author}
  {\bibfnamefont {C.}~\bibnamefont {Chapelier}},\ and\ \bibinfo {author}
  {\bibfnamefont {V.~T.}\ \bibnamefont {Renard}},\ }\bibfield  {title}
  {\bibinfo {title} {Giant {{Atomic Swirl}} in {{Graphene Bilayers}} with
  {{Biaxial Heterostrain}}},\ }\href {https://doi.org/10.1002/adma.202306312}
  {\bibfield  {journal} {\bibinfo  {journal} {Advanced Materials}\ }\textbf
  {\bibinfo {volume} {35}},\ \bibinfo {pages} {2306312} (\bibinfo {year}
  {2023})}\BibitemShut {NoStop}%
\bibitem [{\citenamefont {Kaliteevski}\ \emph {et~al.}(2023)\citenamefont
  {Kaliteevski}, \citenamefont {Enaldiev},\ and\ \citenamefont
  {Fal'ko}}]{kaliteevski2023twirling}%
  \BibitemOpen
  \bibfield  {author} {\bibinfo {author} {\bibfnamefont {M.~A.}\ \bibnamefont
  {Kaliteevski}}, \bibinfo {author} {\bibfnamefont {V.}~\bibnamefont
  {Enaldiev}},\ and\ \bibinfo {author} {\bibfnamefont {V.~I.}\ \bibnamefont
  {Fal'ko}},\ }\bibfield  {title} {\bibinfo {title} {Twirling and {{Spontaneous
  Symmetry Breaking}} of {{Domain Wall Networks}} in {{Lattice-Reconstructed
  Heterostructures}} of {{Two-Dimensional Materials}}},\ }\href
  {https://doi.org/10.1021/acs.nanolett.3c01896} {\bibfield  {journal}
  {\bibinfo  {journal} {Nano Lett.}\ }\textbf {\bibinfo {volume} {23}},\
  \bibinfo {pages} {8875} (\bibinfo {year} {2023})}\BibitemShut {NoStop}%
\bibitem [{\citenamefont {De~Jong}\ \emph {et~al.}(2023)\citenamefont
  {De~Jong}, \citenamefont {Visser}, \citenamefont {Jobst}, \citenamefont
  {Tromp},\ and\ \citenamefont {Van Der~Molen}}]{dejong2023stacking}%
  \BibitemOpen
  \bibfield  {author} {\bibinfo {author} {\bibfnamefont {T.~A.}\ \bibnamefont
  {De~Jong}}, \bibinfo {author} {\bibfnamefont {L.}~\bibnamefont {Visser}},
  \bibinfo {author} {\bibfnamefont {J.}~\bibnamefont {Jobst}}, \bibinfo
  {author} {\bibfnamefont {R.~M.}\ \bibnamefont {Tromp}},\ and\ \bibinfo
  {author} {\bibfnamefont {S.~J.}\ \bibnamefont {Van Der~Molen}},\ }\bibfield
  {title} {\bibinfo {title} {Stacking domain morphology in epitaxial graphene
  on silicon carbide},\ }\href
  {https://doi.org/10.1103/PhysRevMaterials.7.034001} {\bibfield  {journal}
  {\bibinfo  {journal} {Phys. Rev. Materials}\ }\textbf {\bibinfo {volume}
  {7}},\ \bibinfo {pages} {034001} (\bibinfo {year} {2023})}\BibitemShut
  {NoStop}%
\bibitem [{\citenamefont {Jin}\ \emph {et~al.}(2020)\citenamefont {Jin},
  \citenamefont {Olsen}, \citenamefont {Luber},\ and\ \citenamefont
  {Buriak}}]{jin2020van}%
  \BibitemOpen
  \bibfield  {author} {\bibinfo {author} {\bibfnamefont {C.}~\bibnamefont
  {Jin}}, \bibinfo {author} {\bibfnamefont {B.~C.}\ \bibnamefont {Olsen}},
  \bibinfo {author} {\bibfnamefont {E.~J.}\ \bibnamefont {Luber}},\ and\
  \bibinfo {author} {\bibfnamefont {J.~M.}\ \bibnamefont {Buriak}},\ }\bibfield
   {title} {\bibinfo {title} {Van der {{Waals Epitaxy}} of {{Soft Twisted
  Bilayers}}: {{Lattice Relaxation}} and {{Mass Density Waves}}},\ }\href
  {https://doi.org/10.1021/acsnano.0c05310} {\bibfield  {journal} {\bibinfo
  {journal} {ACS Nano}\ }\textbf {\bibinfo {volume} {14}},\ \bibinfo {pages}
  {13441} (\bibinfo {year} {2020})}\BibitemShut {NoStop}%
\bibitem [{\citenamefont {Soltero}\ \emph {et~al.}(2024)\citenamefont
  {Soltero}, \citenamefont {Kaliteevski}, \citenamefont {McHugh}, \citenamefont
  {Enaldiev},\ and\ \citenamefont {Fal'ko}}]{soltero2024competition}%
  \BibitemOpen
  \bibfield  {author} {\bibinfo {author} {\bibfnamefont {I.}~\bibnamefont
  {Soltero}}, \bibinfo {author} {\bibfnamefont {M.~A.}\ \bibnamefont
  {Kaliteevski}}, \bibinfo {author} {\bibfnamefont {J.~G.}\ \bibnamefont
  {McHugh}}, \bibinfo {author} {\bibfnamefont {V.}~\bibnamefont {Enaldiev}},\
  and\ \bibinfo {author} {\bibfnamefont {V.~I.}\ \bibnamefont {Fal'ko}},\
  }\bibfield  {title} {\bibinfo {title} {Competition of {{Moir{\'e} Network
  Sites}} to {{Form Electronic Quantum Dots}} in {{Reconstructed
  MoX}}{\textsubscript{2}} /{{WX}}{\textsubscript{2}} {{Heterostructures}}},\
  }\href {https://doi.org/10.1021/acs.nanolett.3c04427} {\bibfield  {journal}
  {\bibinfo  {journal} {Nano Lett.}\ }\textbf {\bibinfo {volume} {24}},\
  \bibinfo {pages} {1996} (\bibinfo {year} {2024})}\BibitemShut {NoStop}%
\bibitem [{\citenamefont {{San-Jose}}\ \emph {et~al.}(2014)\citenamefont
  {{San-Jose}}, \citenamefont {{Guti{\'e}rrez-Rubio}}, \citenamefont {Sturla},\
  and\ \citenamefont {Guinea}}]{san-jose2014spontaneous}%
  \BibitemOpen
  \bibfield  {author} {\bibinfo {author} {\bibfnamefont {P.}~\bibnamefont
  {{San-Jose}}}, \bibinfo {author} {\bibfnamefont {A.}~\bibnamefont
  {{Guti{\'e}rrez-Rubio}}}, \bibinfo {author} {\bibfnamefont {M.}~\bibnamefont
  {Sturla}},\ and\ \bibinfo {author} {\bibfnamefont {F.}~\bibnamefont
  {Guinea}},\ }\bibfield  {title} {\bibinfo {title} {Spontaneous strains and
  gap in graphene on boron nitride},\ }\href
  {https://doi.org/10.1103/PhysRevB.90.075428} {\bibfield  {journal} {\bibinfo
  {journal} {Phys. Rev. B}\ }\textbf {\bibinfo {volume} {90}},\ \bibinfo
  {pages} {075428} (\bibinfo {year} {2014})}\BibitemShut {NoStop}%
\bibitem [{\citenamefont {Nam}\ and\ \citenamefont
  {Koshino}(2017)}]{nam2017lattice}%
  \BibitemOpen
  \bibfield  {author} {\bibinfo {author} {\bibfnamefont {N.~N.~T.}\
  \bibnamefont {Nam}}\ and\ \bibinfo {author} {\bibfnamefont {M.}~\bibnamefont
  {Koshino}},\ }\bibfield  {title} {\bibinfo {title} {Lattice relaxation and
  energy band modulation in twisted bilayer graphene},\ }\href
  {https://doi.org/10.1103/PhysRevB.96.075311} {\bibfield  {journal} {\bibinfo
  {journal} {Phys. Rev. B}\ }\textbf {\bibinfo {volume} {96}},\ \bibinfo
  {pages} {075311} (\bibinfo {year} {2017})}\BibitemShut {NoStop}%
\bibitem [{\citenamefont {Bennett}(2022)}]{bennett2022theory}%
  \BibitemOpen
  \bibfield  {author} {\bibinfo {author} {\bibfnamefont {D.}~\bibnamefont
  {Bennett}},\ }\bibfield  {title} {\bibinfo {title} {Theory of polar domains
  in moir{\'e} heterostructures},\ }\href
  {https://doi.org/10.1103/PhysRevB.105.235445} {\bibfield  {journal} {\bibinfo
   {journal} {Phys. Rev. B}\ }\textbf {\bibinfo {volume} {105}},\ \bibinfo
  {pages} {235445} (\bibinfo {year} {2022})}\BibitemShut {NoStop}%
\bibitem [{\citenamefont {Koshino}\ and\ \citenamefont
  {Son}(2019)}]{koshino2019moire}%
  \BibitemOpen
  \bibfield  {author} {\bibinfo {author} {\bibfnamefont {M.}~\bibnamefont
  {Koshino}}\ and\ \bibinfo {author} {\bibfnamefont {Y.-W.}\ \bibnamefont
  {Son}},\ }\bibfield  {title} {\bibinfo {title} {Moir{\'e} phonons in twisted
  bilayer graphene},\ }\href {https://doi.org/10.1103/PhysRevB.100.075416}
  {\bibfield  {journal} {\bibinfo  {journal} {Phys. Rev. B}\ }\textbf {\bibinfo
  {volume} {100}},\ \bibinfo {pages} {075416} (\bibinfo {year}
  {2019})}\BibitemShut {NoStop}%
\bibitem [{sup()}]{supplemental}%
  \BibitemOpen
  \href@noop {} {\bibinfo {title} {Supplemental {{Material}}}}\BibitemShut
  {NoStop}%
\bibitem [{Note1()}]{Note1}%
  \BibitemOpen
  \bibinfo {note} {Throughout this paper we refer to the honeycomb lattice
  moir\'e with triangular-lattice arrangement of domain wall junctions as the
  ``triangular system''.}\BibitemShut {Stop}%
\bibitem [{Note2()}]{Note2}%
  \BibitemOpen
  \bibinfo {note} {Negative $V_1$ leads to hexagonal domains, where the domain
  wall junctions are arranged in hexagonal rather than triangular form.
  Frustration does not occur in this case. In square lattices, $V_1$ can be
  converted to $-V_1$ by a redefinition ${\protect \bm {d}}\rightarrow
  {\protect \bm {d}}+ (a/2, \protect \,a/2)$.}\BibitemShut {Stop}%
\bibitem [{\citenamefont {Haastrup}\ \emph {et~al.}(2018)\citenamefont
  {Haastrup}, \citenamefont {Strange}, \citenamefont {Pandey}, \citenamefont
  {Deilmann}, \citenamefont {Schmidt}, \citenamefont {Hinsche}, \citenamefont
  {Gjerding}, \citenamefont {Torelli}, \citenamefont {Larsen}, \citenamefont
  {{Riis-Jensen}}, \citenamefont {Gath}, \citenamefont {Jacobsen},
  \citenamefont {J{\o}rgen~Mortensen}, \citenamefont {Olsen},\ and\
  \citenamefont {Thygesen}}]{haastrup2018computational}%
  \BibitemOpen
  \bibfield  {author} {\bibinfo {author} {\bibfnamefont {S.}~\bibnamefont
  {Haastrup}}, \bibinfo {author} {\bibfnamefont {M.}~\bibnamefont {Strange}},
  \bibinfo {author} {\bibfnamefont {M.}~\bibnamefont {Pandey}}, \bibinfo
  {author} {\bibfnamefont {T.}~\bibnamefont {Deilmann}}, \bibinfo {author}
  {\bibfnamefont {P.~S.}\ \bibnamefont {Schmidt}}, \bibinfo {author}
  {\bibfnamefont {N.~F.}\ \bibnamefont {Hinsche}}, \bibinfo {author}
  {\bibfnamefont {M.~N.}\ \bibnamefont {Gjerding}}, \bibinfo {author}
  {\bibfnamefont {D.}~\bibnamefont {Torelli}}, \bibinfo {author} {\bibfnamefont
  {P.~M.}\ \bibnamefont {Larsen}}, \bibinfo {author} {\bibfnamefont {A.~C.}\
  \bibnamefont {{Riis-Jensen}}}, \bibinfo {author} {\bibfnamefont
  {J.}~\bibnamefont {Gath}}, \bibinfo {author} {\bibfnamefont {K.~W.}\
  \bibnamefont {Jacobsen}}, \bibinfo {author} {\bibfnamefont {J.}~\bibnamefont
  {J{\o}rgen~Mortensen}}, \bibinfo {author} {\bibfnamefont {T.}~\bibnamefont
  {Olsen}},\ and\ \bibinfo {author} {\bibfnamefont {K.~S.}\ \bibnamefont
  {Thygesen}},\ }\bibfield  {title} {\bibinfo {title} {The {{Computational 2D
  Materials Database}}: {{High-throughput}} modeling and discovery of
  atomically thin crystals},\ }\href {https://doi.org/10.1088/2053-1583/aacfc1}
  {\bibfield  {journal} {\bibinfo  {journal} {2D Mater.}\ }\textbf {\bibinfo
  {volume} {5}},\ \bibinfo {pages} {042002} (\bibinfo {year}
  {2018})}\BibitemShut {NoStop}%
\bibitem [{\citenamefont {Gjerding}\ \emph {et~al.}(2021)\citenamefont
  {Gjerding}, \citenamefont {Taghizadeh}, \citenamefont {Rasmussen},
  \citenamefont {Ali}, \citenamefont {Bertoldo}, \citenamefont {Deilmann},
  \citenamefont {Kn{\o}sgaard}, \citenamefont {Kruse}, \citenamefont {Larsen},
  \citenamefont {Manti}, \citenamefont {Pedersen}, \citenamefont {Petralanda},
  \citenamefont {Skovhus}, \citenamefont {Svendsen}, \citenamefont {Mortensen},
  \citenamefont {Olsen},\ and\ \citenamefont {Thygesen}}]{gjerding2021recent}%
  \BibitemOpen
  \bibfield  {author} {\bibinfo {author} {\bibfnamefont {M.~N.}\ \bibnamefont
  {Gjerding}}, \bibinfo {author} {\bibfnamefont {A.}~\bibnamefont
  {Taghizadeh}}, \bibinfo {author} {\bibfnamefont {A.}~\bibnamefont
  {Rasmussen}}, \bibinfo {author} {\bibfnamefont {S.}~\bibnamefont {Ali}},
  \bibinfo {author} {\bibfnamefont {F.}~\bibnamefont {Bertoldo}}, \bibinfo
  {author} {\bibfnamefont {T.}~\bibnamefont {Deilmann}}, \bibinfo {author}
  {\bibfnamefont {N.~R.}\ \bibnamefont {Kn{\o}sgaard}}, \bibinfo {author}
  {\bibfnamefont {M.}~\bibnamefont {Kruse}}, \bibinfo {author} {\bibfnamefont
  {A.~H.}\ \bibnamefont {Larsen}}, \bibinfo {author} {\bibfnamefont
  {S.}~\bibnamefont {Manti}}, \bibinfo {author} {\bibfnamefont {T.~G.}\
  \bibnamefont {Pedersen}}, \bibinfo {author} {\bibfnamefont {U.}~\bibnamefont
  {Petralanda}}, \bibinfo {author} {\bibfnamefont {T.}~\bibnamefont {Skovhus}},
  \bibinfo {author} {\bibfnamefont {M.~K.}\ \bibnamefont {Svendsen}}, \bibinfo
  {author} {\bibfnamefont {J.~J.}\ \bibnamefont {Mortensen}}, \bibinfo {author}
  {\bibfnamefont {T.}~\bibnamefont {Olsen}},\ and\ \bibinfo {author}
  {\bibfnamefont {K.~S.}\ \bibnamefont {Thygesen}},\ }\bibfield  {title}
  {\bibinfo {title} {Recent progress of the {{Computational 2D Materials
  Database}} ({{C2DB}})},\ }\href {https://doi.org/10.1088/2053-1583/ac1059}
  {\bibfield  {journal} {\bibinfo  {journal} {2D Mater.}\ }\textbf {\bibinfo
  {volume} {8}},\ \bibinfo {pages} {044002} (\bibinfo {year}
  {2021})}\BibitemShut {NoStop}%
\bibitem [{\citenamefont {Sachs}\ \emph {et~al.}(2011)\citenamefont {Sachs},
  \citenamefont {Wehling}, \citenamefont {Katsnelson},\ and\ \citenamefont
  {Lichtenstein}}]{sachs2011adhesion}%
  \BibitemOpen
  \bibfield  {author} {\bibinfo {author} {\bibfnamefont {B.}~\bibnamefont
  {Sachs}}, \bibinfo {author} {\bibfnamefont {T.~O.}\ \bibnamefont {Wehling}},
  \bibinfo {author} {\bibfnamefont {M.~I.}\ \bibnamefont {Katsnelson}},\ and\
  \bibinfo {author} {\bibfnamefont {A.~I.}\ \bibnamefont {Lichtenstein}},\
  }\bibfield  {title} {\bibinfo {title} {Adhesion and electronic structure of
  graphene on hexagonal boron nitride substrates},\ }\href
  {https://doi.org/10.1103/PhysRevB.84.195414} {\bibfield  {journal} {\bibinfo
  {journal} {Phys. Rev. B}\ }\textbf {\bibinfo {volume} {84}},\ \bibinfo
  {pages} {195414} (\bibinfo {year} {2011})}\BibitemShut {NoStop}%
\bibitem [{\citenamefont {Wannier}(1950)}]{wannier1950antiferromagnetism}%
  \BibitemOpen
  \bibfield  {author} {\bibinfo {author} {\bibfnamefont {G.~H.}\ \bibnamefont
  {Wannier}},\ }\bibfield  {title} {\bibinfo {title} {Antiferromagnetism.
  {{The}} triangular ising net},\ }\href
  {https://doi.org/10.1103/PhysRev.79.357} {\bibfield  {journal} {\bibinfo
  {journal} {Phys. Rev.}\ }\textbf {\bibinfo {volume} {79}},\ \bibinfo {pages}
  {357} (\bibinfo {year} {1950})}\BibitemShut {NoStop}%
\bibitem [{\citenamefont {Maignan}\ \emph {et~al.}(2000)\citenamefont
  {Maignan}, \citenamefont {Michel}, \citenamefont {Masset}, \citenamefont
  {Martin},\ and\ \citenamefont {Raveau}}]{maignan2000single}%
  \BibitemOpen
  \bibfield  {author} {\bibinfo {author} {\bibfnamefont {A.}~\bibnamefont
  {Maignan}}, \bibinfo {author} {\bibfnamefont {C.}~\bibnamefont {Michel}},
  \bibinfo {author} {\bibfnamefont {A.}~\bibnamefont {Masset}}, \bibinfo
  {author} {\bibfnamefont {C.}~\bibnamefont {Martin}},\ and\ \bibinfo {author}
  {\bibfnamefont {B.}~\bibnamefont {Raveau}},\ }\bibfield  {title} {\bibinfo
  {title} {Single crystal study of the one dimensional {{Ca Co O}} compound:
  Five stable configurations for the {{Ising}} triangular lattice},\ }\href
  {https://doi.org/10.1007/pl00011051} {\bibfield  {journal} {\bibinfo
  {journal} {Eur. Phys. J. B}\ }\textbf {\bibinfo {volume} {15}},\ \bibinfo
  {pages} {657} (\bibinfo {year} {2000})}\BibitemShut {NoStop}%
\bibitem [{\citenamefont {Hardy}\ \emph {et~al.}(2004)\citenamefont {Hardy},
  \citenamefont {Lees}, \citenamefont {Petrenko}, \citenamefont {Paul},
  \citenamefont {Flahaut}, \citenamefont {H{\'e}bert},\ and\ \citenamefont
  {Maignan}}]{hardy2004temperature}%
  \BibitemOpen
  \bibfield  {author} {\bibinfo {author} {\bibfnamefont {V.}~\bibnamefont
  {Hardy}}, \bibinfo {author} {\bibfnamefont {M.~R.}\ \bibnamefont {Lees}},
  \bibinfo {author} {\bibfnamefont {O.~A.}\ \bibnamefont {Petrenko}}, \bibinfo
  {author} {\bibfnamefont {D.~{\relax McK}.}\ \bibnamefont {Paul}}, \bibinfo
  {author} {\bibfnamefont {D.}~\bibnamefont {Flahaut}}, \bibinfo {author}
  {\bibfnamefont {S.}~\bibnamefont {H{\'e}bert}},\ and\ \bibinfo {author}
  {\bibfnamefont {A.}~\bibnamefont {Maignan}},\ }\bibfield  {title} {\bibinfo
  {title} {Temperature and time dependence of the field-driven magnetization
  steps {{inCa3Co2O6single}} crystals},\ }\href
  {https://doi.org/10.1103/physrevb.70.064424} {\bibfield  {journal} {\bibinfo
  {journal} {Phys. Rev. B}\ }\textbf {\bibinfo {volume} {70}},\ \bibinfo
  {pages} {064424} (\bibinfo {year} {2004})}\BibitemShut {NoStop}%
\bibitem [{Note3()}]{Note3}%
  \BibitemOpen
  \bibinfo {note} {Note the subtlety that the size and orientation of moir\'e
  cell changes with the twist angle. Iterating from the exact same ${\protect
  \bm {\delta }}({\protect \bm {r}})$ at each ${\protect \bm {r}}$ point of the
  previous sample would suffer difficulties due to incompatibility in
  superlattice periodicity. Therefore, at each inter-sample transition, we
  scale our spatial grid to fit the next supercell without changing the
  magnitudes or the directions of ${\protect \bm {\delta }}({\protect \bm
  {r}})$'s on the grid points. This is expected to be a good approximation in
  the adiabatic limit.}\BibitemShut {Stop}%
\bibitem [{\citenamefont {Kageyama}\ \emph {et~al.}(1997)\citenamefont
  {Kageyama}, \citenamefont {Yoshimura}, \citenamefont {Kosuge}, \citenamefont
  {Mitamura},\ and\ \citenamefont {Goto}}]{kageyama1997fieldinduced}%
  \BibitemOpen
  \bibfield  {author} {\bibinfo {author} {\bibfnamefont {H.}~\bibnamefont
  {Kageyama}}, \bibinfo {author} {\bibfnamefont {K.}~\bibnamefont {Yoshimura}},
  \bibinfo {author} {\bibfnamefont {K.}~\bibnamefont {Kosuge}}, \bibinfo
  {author} {\bibfnamefont {H.}~\bibnamefont {Mitamura}},\ and\ \bibinfo
  {author} {\bibfnamefont {T.}~\bibnamefont {Goto}},\ }\bibfield  {title}
  {\bibinfo {title} {Field-{{Induced Magnetic Transitions}} in the
  {{One-Dimensional Compound
  Ca}}{\textsubscript{{\textbf{3}}}}{{Co}}{\textsubscript{{\textbf{2}}}}{{O}}{\textsubscript{{\textbf{6}}}}},\
  }\href {https://doi.org/10.1143/jpsj.66.1607} {\bibfield  {journal} {\bibinfo
   {journal} {J. Phys. Soc. Jpn.}\ }\textbf {\bibinfo {volume} {66}},\ \bibinfo
  {pages} {1607} (\bibinfo {year} {1997})}\BibitemShut {NoStop}%
\bibitem [{\citenamefont {Kudasov}(2006)}]{kudasov2006steplike}%
  \BibitemOpen
  \bibfield  {author} {\bibinfo {author} {\bibfnamefont {Y.~B.}\ \bibnamefont
  {Kudasov}},\ }\bibfield  {title} {\bibinfo {title} {Steplike
  {{Magnetization}} in a {{Spin-Chain System}}:{{Ca3Co2O6}}},\ }\href
  {https://doi.org/10.1103/physrevlett.96.027212} {\bibfield  {journal}
  {\bibinfo  {journal} {Phys. Rev. Lett.}\ }\textbf {\bibinfo {volume} {96}},\
  \bibinfo {pages} {027212} (\bibinfo {year} {2006})}\BibitemShut {NoStop}%
\bibitem [{\citenamefont {Kudasov}\ \emph {et~al.}(2008)\citenamefont
  {Kudasov}, \citenamefont {Korshunov}, \citenamefont {Pavlov},\ and\
  \citenamefont {Maslov}}]{kudasov2008dynamics}%
  \BibitemOpen
  \bibfield  {author} {\bibinfo {author} {\bibfnamefont {{\relax Yu}.~B.}\
  \bibnamefont {Kudasov}}, \bibinfo {author} {\bibfnamefont {A.~S.}\
  \bibnamefont {Korshunov}}, \bibinfo {author} {\bibfnamefont {V.~N.}\
  \bibnamefont {Pavlov}},\ and\ \bibinfo {author} {\bibfnamefont {D.~A.}\
  \bibnamefont {Maslov}},\ }\bibfield  {title} {\bibinfo {title} {Dynamics of
  magnetization in frustrated spin-chain {{systemCa3Co2O6}}},\ }\href
  {https://doi.org/10.1103/physrevb.78.132407} {\bibfield  {journal} {\bibinfo
  {journal} {Phys. Rev. B}\ }\textbf {\bibinfo {volume} {78}},\ \bibinfo
  {pages} {132407} (\bibinfo {year} {2008})}\BibitemShut {NoStop}%
\bibitem [{\citenamefont {Yao}\ \emph {et~al.}(2006)\citenamefont {Yao},
  \citenamefont {Dong}, \citenamefont {Yu},\ and\ \citenamefont
  {Liu}}]{yao2006monte}%
  \BibitemOpen
  \bibfield  {author} {\bibinfo {author} {\bibfnamefont {X.}~\bibnamefont
  {Yao}}, \bibinfo {author} {\bibfnamefont {S.}~\bibnamefont {Dong}}, \bibinfo
  {author} {\bibfnamefont {H.}~\bibnamefont {Yu}},\ and\ \bibinfo {author}
  {\bibfnamefont {J.}~\bibnamefont {Liu}},\ }\bibfield  {title} {\bibinfo
  {title} {Monte {{Carlo}} simulation of magnetic behavior of a spin-chain
  system on a triangular lattice},\ }\href
  {https://doi.org/10.1103/physrevb.74.134421} {\bibfield  {journal} {\bibinfo
  {journal} {Phys. Rev. B}\ }\textbf {\bibinfo {volume} {74}},\ \bibinfo
  {pages} {134421} (\bibinfo {year} {2006})}\BibitemShut {NoStop}%
\bibitem [{\citenamefont {{Ribeiro-Palau}}\ \emph {et~al.}(2018)\citenamefont
  {{Ribeiro-Palau}}, \citenamefont {Zhang}, \citenamefont {Watanabe},
  \citenamefont {Taniguchi}, \citenamefont {Hone},\ and\ \citenamefont
  {Dean}}]{ribeiro-palau2018twistable}%
  \BibitemOpen
  \bibfield  {author} {\bibinfo {author} {\bibfnamefont {R.}~\bibnamefont
  {{Ribeiro-Palau}}}, \bibinfo {author} {\bibfnamefont {C.}~\bibnamefont
  {Zhang}}, \bibinfo {author} {\bibfnamefont {K.}~\bibnamefont {Watanabe}},
  \bibinfo {author} {\bibfnamefont {T.}~\bibnamefont {Taniguchi}}, \bibinfo
  {author} {\bibfnamefont {J.}~\bibnamefont {Hone}},\ and\ \bibinfo {author}
  {\bibfnamefont {C.~R.}\ \bibnamefont {Dean}},\ }\bibfield  {title} {\bibinfo
  {title} {Twistable electronics with dynamically rotatable heterostructures},\
  }\href {https://doi.org/10.1126/science.aat6981} {\bibfield  {journal}
  {\bibinfo  {journal} {Science}\ }\textbf {\bibinfo {volume} {361}},\ \bibinfo
  {pages} {690} (\bibinfo {year} {2018})}\BibitemShut {NoStop}%
\bibitem [{\citenamefont {Inbar}\ \emph {et~al.}(2023)\citenamefont {Inbar},
  \citenamefont {Birkbeck}, \citenamefont {Xiao}, \citenamefont {Taniguchi},
  \citenamefont {Watanabe}, \citenamefont {Yan}, \citenamefont {Oreg},
  \citenamefont {Stern}, \citenamefont {Berg},\ and\ \citenamefont
  {Ilani}}]{inbar2023quantum}%
  \BibitemOpen
  \bibfield  {author} {\bibinfo {author} {\bibfnamefont {A.}~\bibnamefont
  {Inbar}}, \bibinfo {author} {\bibfnamefont {J.}~\bibnamefont {Birkbeck}},
  \bibinfo {author} {\bibfnamefont {J.}~\bibnamefont {Xiao}}, \bibinfo {author}
  {\bibfnamefont {T.}~\bibnamefont {Taniguchi}}, \bibinfo {author}
  {\bibfnamefont {K.}~\bibnamefont {Watanabe}}, \bibinfo {author}
  {\bibfnamefont {B.}~\bibnamefont {Yan}}, \bibinfo {author} {\bibfnamefont
  {Y.}~\bibnamefont {Oreg}}, \bibinfo {author} {\bibfnamefont {A.}~\bibnamefont
  {Stern}}, \bibinfo {author} {\bibfnamefont {E.}~\bibnamefont {Berg}},\ and\
  \bibinfo {author} {\bibfnamefont {S.}~\bibnamefont {Ilani}},\ }\bibfield
  {title} {\bibinfo {title} {The quantum twisting microscope},\ }\href
  {https://doi.org/10.1038/s41586-022-05685-y} {\bibfield  {journal} {\bibinfo
  {journal} {Nature}\ }\textbf {\bibinfo {volume} {614}},\ \bibinfo {pages}
  {682} (\bibinfo {year} {2023})}\BibitemShut {NoStop}%
\bibitem [{\citenamefont {Farrar}\ \emph {et~al.}(2025)\citenamefont {Farrar},
  \citenamefont {Maffione}, \citenamefont {Nguyen}, \citenamefont {Watanabe},
  \citenamefont {Taniguchi}, \citenamefont {Charlier}, \citenamefont {Mailly},\
  and\ \citenamefont {{Ribeiro-Palau}}}]{farrar2025impact}%
  \BibitemOpen
  \bibfield  {author} {\bibinfo {author} {\bibfnamefont {L.~S.}\ \bibnamefont
  {Farrar}}, \bibinfo {author} {\bibfnamefont {G.}~\bibnamefont {Maffione}},
  \bibinfo {author} {\bibfnamefont {V.-H.}\ \bibnamefont {Nguyen}}, \bibinfo
  {author} {\bibfnamefont {K.}~\bibnamefont {Watanabe}}, \bibinfo {author}
  {\bibfnamefont {T.}~\bibnamefont {Taniguchi}}, \bibinfo {author}
  {\bibfnamefont {J.-C.}\ \bibnamefont {Charlier}}, \bibinfo {author}
  {\bibfnamefont {D.}~\bibnamefont {Mailly}},\ and\ \bibinfo {author}
  {\bibfnamefont {R.}~\bibnamefont {{Ribeiro-Palau}}},\ }\bibfield  {title}
  {\bibinfo {title} {Impact of the {{Angular Alignment}} on the {{Crystal
  Field}} and {{Intrinsic Doping}} of {{Bilayer Graphene}}/{{BN
  Heterostructures}}},\ }\href {https://doi.org/10.1021/acs.nanolett.4c05378}
  {\bibfield  {journal} {\bibinfo  {journal} {Nano Lett.}\ }\textbf {\bibinfo
  {volume} {25}},\ \bibinfo {pages} {2236} (\bibinfo {year}
  {2025})}\BibitemShut {NoStop}%
\bibitem [{\citenamefont {Anderson}(1958)}]{anderson1958absence}%
  \BibitemOpen
  \bibfield  {author} {\bibinfo {author} {\bibfnamefont {P.~W.}\ \bibnamefont
  {Anderson}},\ }\bibfield  {title} {\bibinfo {title} {Absence of {{Diffusion}}
  in {{Certain Random Lattices}}},\ }\href
  {https://doi.org/10.1103/PhysRev.109.1492} {\bibfield  {journal} {\bibinfo
  {journal} {Phys. Rev.}\ }\textbf {\bibinfo {volume} {109}},\ \bibinfo {pages}
  {1492} (\bibinfo {year} {1958})}\BibitemShut {NoStop}%
\bibitem [{\citenamefont {Yasuda}\ \emph {et~al.}(2021)\citenamefont {Yasuda},
  \citenamefont {Wang}, \citenamefont {Watanabe}, \citenamefont {Taniguchi},\
  and\ \citenamefont {{Jarillo-Herrero}}}]{yasuda2021stackingengineered}%
  \BibitemOpen
  \bibfield  {author} {\bibinfo {author} {\bibfnamefont {K.}~\bibnamefont
  {Yasuda}}, \bibinfo {author} {\bibfnamefont {X.}~\bibnamefont {Wang}},
  \bibinfo {author} {\bibfnamefont {K.}~\bibnamefont {Watanabe}}, \bibinfo
  {author} {\bibfnamefont {T.}~\bibnamefont {Taniguchi}},\ and\ \bibinfo
  {author} {\bibfnamefont {P.}~\bibnamefont {{Jarillo-Herrero}}},\ }\bibfield
  {title} {\bibinfo {title} {Stacking-engineered ferroelectricity in bilayer
  boron nitride},\ }\href {https://doi.org/10.1126/science.abd323} {\bibfield
  {journal} {\bibinfo  {journal} {Science}\ }\textbf {\bibinfo {volume}
  {372}},\ \bibinfo {pages} {1458} (\bibinfo {year} {2021})}\BibitemShut
  {NoStop}%
\bibitem [{\citenamefont {Zheng}\ \emph {et~al.}(2020)\citenamefont {Zheng},
  \citenamefont {Ma}, \citenamefont {Bi}, \citenamefont {{de La Barrera}},
  \citenamefont {Liu}, \citenamefont {Mao}, \citenamefont {Zhang},
  \citenamefont {Kiper}, \citenamefont {Watanabe}, \citenamefont {Taniguchi}
  \emph {et~al.}}]{zheng2020unconventional}%
  \BibitemOpen
  \bibfield  {author} {\bibinfo {author} {\bibfnamefont {Z.}~\bibnamefont
  {Zheng}}, \bibinfo {author} {\bibfnamefont {Q.}~\bibnamefont {Ma}}, \bibinfo
  {author} {\bibfnamefont {Z.}~\bibnamefont {Bi}}, \bibinfo {author}
  {\bibfnamefont {S.}~\bibnamefont {{de La Barrera}}}, \bibinfo {author}
  {\bibfnamefont {M.-H.}\ \bibnamefont {Liu}}, \bibinfo {author} {\bibfnamefont
  {N.}~\bibnamefont {Mao}}, \bibinfo {author} {\bibfnamefont {Y.}~\bibnamefont
  {Zhang}}, \bibinfo {author} {\bibfnamefont {N.}~\bibnamefont {Kiper}},
  \bibinfo {author} {\bibfnamefont {K.}~\bibnamefont {Watanabe}}, \bibinfo
  {author} {\bibfnamefont {T.}~\bibnamefont {Taniguchi}}, \emph {et~al.},\
  }\bibfield  {title} {\bibinfo {title} {Unconventional ferroelectricity in
  moir{\'e} heterostructures},\ }\href
  {https://doi.org/10.1038/s41586-020-2970-9} {\bibfield  {journal} {\bibinfo
  {journal} {Nature}\ }\textbf {\bibinfo {volume} {588}},\ \bibinfo {pages}
  {71} (\bibinfo {year} {2020})}\BibitemShut {NoStop}%
\bibitem [{\citenamefont {Niu}\ \emph {et~al.}(2022)\citenamefont {Niu},
  \citenamefont {Li}, \citenamefont {Han}, \citenamefont {Qu}, \citenamefont
  {Ding}, \citenamefont {Wang}, \citenamefont {Liu}, \citenamefont {Liu},
  \citenamefont {Han}, \citenamefont {Watanabe} \emph {et~al.}}]{niu2022giant}%
  \BibitemOpen
  \bibfield  {author} {\bibinfo {author} {\bibfnamefont {R.}~\bibnamefont
  {Niu}}, \bibinfo {author} {\bibfnamefont {Z.}~\bibnamefont {Li}}, \bibinfo
  {author} {\bibfnamefont {X.}~\bibnamefont {Han}}, \bibinfo {author}
  {\bibfnamefont {Z.}~\bibnamefont {Qu}}, \bibinfo {author} {\bibfnamefont
  {D.}~\bibnamefont {Ding}}, \bibinfo {author} {\bibfnamefont {Z.}~\bibnamefont
  {Wang}}, \bibinfo {author} {\bibfnamefont {Q.}~\bibnamefont {Liu}}, \bibinfo
  {author} {\bibfnamefont {T.}~\bibnamefont {Liu}}, \bibinfo {author}
  {\bibfnamefont {C.}~\bibnamefont {Han}}, \bibinfo {author} {\bibfnamefont
  {K.}~\bibnamefont {Watanabe}}, \emph {et~al.},\ }\bibfield  {title} {\bibinfo
  {title} {Giant ferroelectric polarization in a bilayer graphene
  heterostructure},\ }\href {https://doi.org/10.1038/s41467-022-34104-z}
  {\bibfield  {journal} {\bibinfo  {journal} {Nature Communications}\ }\textbf
  {\bibinfo {volume} {13}},\ \bibinfo {pages} {6241} (\bibinfo {year}
  {2022})}\BibitemShut {NoStop}%
\bibitem [{\citenamefont {Zheng}\ \emph {et~al.}(2023)\citenamefont {Zheng},
  \citenamefont {Wang}, \citenamefont {Zhu}, \citenamefont {Carr},
  \citenamefont {Devakul}, \citenamefont {de~la Barrera}, \citenamefont {Paul},
  \citenamefont {Huang}, \citenamefont {Gao}, \citenamefont {Zhang},
  \citenamefont {B{\'e}rub{\'e}}, \citenamefont {Evancho}, \citenamefont
  {Watanabe}, \citenamefont {Taniguchi}, \citenamefont {Fu}, \citenamefont
  {Wang}, \citenamefont {Xu}, \citenamefont {Kaxiras}, \citenamefont
  {{Jarillo-Herrero}},\ and\ \citenamefont {Ma}}]{zheng2023electronic}%
  \BibitemOpen
  \bibfield  {author} {\bibinfo {author} {\bibfnamefont {Z.}~\bibnamefont
  {Zheng}}, \bibinfo {author} {\bibfnamefont {X.}~\bibnamefont {Wang}},
  \bibinfo {author} {\bibfnamefont {Z.}~\bibnamefont {Zhu}}, \bibinfo {author}
  {\bibfnamefont {S.}~\bibnamefont {Carr}}, \bibinfo {author} {\bibfnamefont
  {T.}~\bibnamefont {Devakul}}, \bibinfo {author} {\bibfnamefont
  {S.}~\bibnamefont {de~la Barrera}}, \bibinfo {author} {\bibfnamefont
  {N.}~\bibnamefont {Paul}}, \bibinfo {author} {\bibfnamefont {Z.}~\bibnamefont
  {Huang}}, \bibinfo {author} {\bibfnamefont {A.}~\bibnamefont {Gao}}, \bibinfo
  {author} {\bibfnamefont {Y.}~\bibnamefont {Zhang}}, \bibinfo {author}
  {\bibfnamefont {D.}~\bibnamefont {B{\'e}rub{\'e}}}, \bibinfo {author}
  {\bibfnamefont {K.~N.}\ \bibnamefont {Evancho}}, \bibinfo {author}
  {\bibfnamefont {K.}~\bibnamefont {Watanabe}}, \bibinfo {author}
  {\bibfnamefont {T.}~\bibnamefont {Taniguchi}}, \bibinfo {author}
  {\bibfnamefont {L.}~\bibnamefont {Fu}}, \bibinfo {author} {\bibfnamefont
  {Y.}~\bibnamefont {Wang}}, \bibinfo {author} {\bibfnamefont {S.-Y.}\
  \bibnamefont {Xu}}, \bibinfo {author} {\bibfnamefont {E.}~\bibnamefont
  {Kaxiras}}, \bibinfo {author} {\bibfnamefont {P.}~\bibnamefont
  {{Jarillo-Herrero}}},\ and\ \bibinfo {author} {\bibfnamefont
  {Q.}~\bibnamefont {Ma}},\ }\href {https://doi.org/10.48550/arXiv.2306.03922}
  {\bibinfo {title} {Electronic ratchet effect in a moir{\'e} system:
  Signatures of excitonic ferroelectricity}} (\bibinfo {year} {2023}),\ \Eprint
  {https://arxiv.org/abs/2306.03922} {arXiv:2306.03922 [cond-mat]} \BibitemShut
  {NoStop}%
\bibitem [{\citenamefont {Zhang}\ \emph {et~al.}(2024)\citenamefont {Zhang},
  \citenamefont {Ding}, \citenamefont {Xiang}, \citenamefont {Liu},
  \citenamefont {Zhou}, \citenamefont {Wu}, \citenamefont {Xin}, \citenamefont
  {Watanabe}, \citenamefont {Taniguchi},\ and\ \citenamefont
  {Xu}}]{zhang2024electronic}%
  \BibitemOpen
  \bibfield  {author} {\bibinfo {author} {\bibfnamefont {L.}~\bibnamefont
  {Zhang}}, \bibinfo {author} {\bibfnamefont {J.}~\bibnamefont {Ding}},
  \bibinfo {author} {\bibfnamefont {H.}~\bibnamefont {Xiang}}, \bibinfo
  {author} {\bibfnamefont {N.}~\bibnamefont {Liu}}, \bibinfo {author}
  {\bibfnamefont {W.}~\bibnamefont {Zhou}}, \bibinfo {author} {\bibfnamefont
  {L.}~\bibnamefont {Wu}}, \bibinfo {author} {\bibfnamefont {N.}~\bibnamefont
  {Xin}}, \bibinfo {author} {\bibfnamefont {K.}~\bibnamefont {Watanabe}},
  \bibinfo {author} {\bibfnamefont {T.}~\bibnamefont {Taniguchi}},\ and\
  \bibinfo {author} {\bibfnamefont {S.}~\bibnamefont {Xu}},\ }\bibfield
  {title} {\bibinfo {title} {Electronic ferroelectricity in monolayer graphene
  moir{\'e} superlattices},\ }\href
  {https://doi.org/10.1038/s41467-024-55281-z} {\bibfield  {journal} {\bibinfo
  {journal} {Nature Communications}\ }\textbf {\bibinfo {volume} {15}},\
  \bibinfo {pages} {10905} (\bibinfo {year} {2024})}\BibitemShut {NoStop}%
\bibitem [{\citenamefont {Yan}\ \emph {et~al.}(2023)\citenamefont {Yan},
  \citenamefont {Zheng}, \citenamefont {Sangwan}, \citenamefont {Qian},
  \citenamefont {Wang}, \citenamefont {Liu}, \citenamefont {Watanabe},
  \citenamefont {Taniguchi}, \citenamefont {Xu}, \citenamefont
  {{Jarillo-Herrero}} \emph {et~al.}}]{yan2023moire}%
  \BibitemOpen
  \bibfield  {author} {\bibinfo {author} {\bibfnamefont {X.}~\bibnamefont
  {Yan}}, \bibinfo {author} {\bibfnamefont {Z.}~\bibnamefont {Zheng}}, \bibinfo
  {author} {\bibfnamefont {V.~K.}\ \bibnamefont {Sangwan}}, \bibinfo {author}
  {\bibfnamefont {J.~H.}\ \bibnamefont {Qian}}, \bibinfo {author}
  {\bibfnamefont {X.}~\bibnamefont {Wang}}, \bibinfo {author} {\bibfnamefont
  {S.~E.}\ \bibnamefont {Liu}}, \bibinfo {author} {\bibfnamefont
  {K.}~\bibnamefont {Watanabe}}, \bibinfo {author} {\bibfnamefont
  {T.}~\bibnamefont {Taniguchi}}, \bibinfo {author} {\bibfnamefont {S.-Y.}\
  \bibnamefont {Xu}}, \bibinfo {author} {\bibfnamefont {P.}~\bibnamefont
  {{Jarillo-Herrero}}}, \emph {et~al.},\ }\bibfield  {title} {\bibinfo {title}
  {Moir{\'e} synaptic transistor with room-temperature neuromorphic
  functionality},\ }\href {https://doi.org/10.1038/s41586-023-06791-1}
  {\bibfield  {journal} {\bibinfo  {journal} {Nature}\ }\textbf {\bibinfo
  {volume} {624}},\ \bibinfo {pages} {551} (\bibinfo {year}
  {2023})}\BibitemShut {NoStop}%
\bibitem [{\citenamefont {Waters}\ \emph {et~al.}(2025)\citenamefont {Waters},
  \citenamefont {Waleffe}, \citenamefont {Thompson}, \citenamefont
  {{Arreguin-Martinez}}, \citenamefont {Fonseca}, \citenamefont {Poirier},
  \citenamefont {Edgar}, \citenamefont {Watanabe}, \citenamefont {Taniguchi},
  \citenamefont {Xu}, \citenamefont {Cobden},\ and\ \citenamefont
  {Yankowitz}}]{waters2025anomalous}%
  \BibitemOpen
  \bibfield  {author} {\bibinfo {author} {\bibfnamefont {D.}~\bibnamefont
  {Waters}}, \bibinfo {author} {\bibfnamefont {D.}~\bibnamefont {Waleffe}},
  \bibinfo {author} {\bibfnamefont {E.}~\bibnamefont {Thompson}}, \bibinfo
  {author} {\bibfnamefont {E.}~\bibnamefont {{Arreguin-Martinez}}}, \bibinfo
  {author} {\bibfnamefont {J.}~\bibnamefont {Fonseca}}, \bibinfo {author}
  {\bibfnamefont {T.}~\bibnamefont {Poirier}}, \bibinfo {author} {\bibfnamefont
  {J.~H.}\ \bibnamefont {Edgar}}, \bibinfo {author} {\bibfnamefont
  {K.}~\bibnamefont {Watanabe}}, \bibinfo {author} {\bibfnamefont
  {T.}~\bibnamefont {Taniguchi}}, \bibinfo {author} {\bibfnamefont
  {X.}~\bibnamefont {Xu}}, \bibinfo {author} {\bibfnamefont {D.}~\bibnamefont
  {Cobden}},\ and\ \bibinfo {author} {\bibfnamefont {M.}~\bibnamefont
  {Yankowitz}},\ }\bibfield  {title} {\bibinfo {title} {Anomalous
  {{Hysteresis}} in {{Graphite}}/{{Boron Nitride Transistors}}},\ }\href
  {https://doi.org/10.1021/acs.nanolett.5c01799} {\bibfield  {journal}
  {\bibinfo  {journal} {Nano Lett.}\ }\textbf {\bibinfo {volume} {25}},\
  \bibinfo {pages} {8768} (\bibinfo {year} {2025})}\BibitemShut {NoStop}%
\bibitem [{\citenamefont {Niu}\ \emph {et~al.}(2025)\citenamefont {Niu},
  \citenamefont {Li}, \citenamefont {Han}, \citenamefont {Qu}, \citenamefont
  {Liu}, \citenamefont {Wang}, \citenamefont {Han}, \citenamefont {Wang},
  \citenamefont {Wu}, \citenamefont {Yang} \emph
  {et~al.}}]{niu2025ferroelectricity}%
  \BibitemOpen
  \bibfield  {author} {\bibinfo {author} {\bibfnamefont {R.}~\bibnamefont
  {Niu}}, \bibinfo {author} {\bibfnamefont {Z.}~\bibnamefont {Li}}, \bibinfo
  {author} {\bibfnamefont {X.}~\bibnamefont {Han}}, \bibinfo {author}
  {\bibfnamefont {Z.}~\bibnamefont {Qu}}, \bibinfo {author} {\bibfnamefont
  {Q.}~\bibnamefont {Liu}}, \bibinfo {author} {\bibfnamefont {Z.}~\bibnamefont
  {Wang}}, \bibinfo {author} {\bibfnamefont {C.}~\bibnamefont {Han}}, \bibinfo
  {author} {\bibfnamefont {C.}~\bibnamefont {Wang}}, \bibinfo {author}
  {\bibfnamefont {Y.}~\bibnamefont {Wu}}, \bibinfo {author} {\bibfnamefont
  {C.}~\bibnamefont {Yang}}, \emph {et~al.},\ }\bibfield  {title} {\bibinfo
  {title} {Ferroelectricity with concomitant {{Coulomb}} screening in van der
  {{Waals}} heterostructures},\ }\href
  {https://doi.org/10.1038/s41565-024-01846-4} {\bibfield  {journal} {\bibinfo
  {journal} {Nat. Nanotechnol.}\ ,\ \bibinfo {pages} {1}} (\bibinfo {year}
  {2025})}\BibitemShut {NoStop}%
\bibitem [{\citenamefont {Press}(2007)}]{press2007numerical}%
  \BibitemOpen
  \bibinfo {editor} {\bibfnamefont {W.~H.}\ \bibnamefont {Press}},\ ed.,\
  \href@noop {} {\emph {\bibinfo {title} {Numerical Recipes: The Art of
  Scientific Computing}}},\ \bibinfo {edition} {3rd}\ ed.\ (\bibinfo
  {publisher} {Cambridge University Press},\ \bibinfo {address} {Cambridge, UK
  ; New York},\ \bibinfo {year} {2007})\BibitemShut {NoStop}%
\bibitem [{\citenamefont {Enaldiev}\ \emph {et~al.}(2022)\citenamefont
  {Enaldiev}, \citenamefont {Ferreira},\ and\ \citenamefont
  {Fal'ko}}]{enaldiev2022scalable}%
  \BibitemOpen
  \bibfield  {author} {\bibinfo {author} {\bibfnamefont {V.~V.}\ \bibnamefont
  {Enaldiev}}, \bibinfo {author} {\bibfnamefont {F.}~\bibnamefont {Ferreira}},\
  and\ \bibinfo {author} {\bibfnamefont {V.~I.}\ \bibnamefont {Fal'ko}},\
  }\bibfield  {title} {\bibinfo {title} {A {{Scalable Network Model}} for
  {{Electrically Tunable Ferroelectric Domain Structure}} in {{Twistronic
  Bilayers}} of {{Two-Dimensional Semiconductors}}},\ }\href
  {https://doi.org/10.1021/acs.nanolett.1c04210} {\bibfield  {journal}
  {\bibinfo  {journal} {Nano Lett.}\ }\textbf {\bibinfo {volume} {22}},\
  \bibinfo {pages} {1534} (\bibinfo {year} {2022})}\BibitemShut {NoStop}%
\bibitem [{\citenamefont {{Ramos-Alonso}}\ \emph {et~al.}(2025)\citenamefont
  {{Ramos-Alonso}}, \citenamefont {Remez}, \citenamefont {Bennett},
  \citenamefont {Fernandes},\ and\ \citenamefont
  {Ochoa}}]{ramos-alonso2025flat}%
  \BibitemOpen
  \bibfield  {author} {\bibinfo {author} {\bibfnamefont {A.}~\bibnamefont
  {{Ramos-Alonso}}}, \bibinfo {author} {\bibfnamefont {B.}~\bibnamefont
  {Remez}}, \bibinfo {author} {\bibfnamefont {D.}~\bibnamefont {Bennett}},
  \bibinfo {author} {\bibfnamefont {R.~M.}\ \bibnamefont {Fernandes}},\ and\
  \bibinfo {author} {\bibfnamefont {H.}~\bibnamefont {Ochoa}},\ }\bibfield
  {title} {\bibinfo {title} {Flat and {{Tunable Moir{\'e} Phonons}} in
  {{Twisted Transition-Metal Dichalcogenides}}},\ }\href
  {https://doi.org/10.1103/PhysRevLett.134.026501} {\bibfield  {journal}
  {\bibinfo  {journal} {Phys. Rev. Lett.}\ }\textbf {\bibinfo {volume} {134}},\
  \bibinfo {pages} {026501} (\bibinfo {year} {2025})}\BibitemShut {NoStop}%
\bibitem [{\citenamefont {Tang}\ and\ \citenamefont
  {Bauer}(2023)}]{tang2023sliding}%
  \BibitemOpen
  \bibfield  {author} {\bibinfo {author} {\bibfnamefont {P.}~\bibnamefont
  {Tang}}\ and\ \bibinfo {author} {\bibfnamefont {G.~E.~W.}\ \bibnamefont
  {Bauer}},\ }\bibfield  {title} {\bibinfo {title} {Sliding {{Phase
  Transition}} in {{Ferroelectric}} van der {{Waals Bilayers}}},\ }\href
  {https://doi.org/10.1103/PhysRevLett.130.176801} {\bibfield  {journal}
  {\bibinfo  {journal} {Phys. Rev. Lett.}\ }\textbf {\bibinfo {volume} {130}},\
  \bibinfo {pages} {176801} (\bibinfo {year} {2023})}\BibitemShut {NoStop}%
\end{thebibliography}%

\clearpage

\appendix
\onecolumngrid
\renewcommand\theequation{S\arabic{equation}}
\renewcommand\thefigure{S\arabic{figure}}
\setcounter{equation}{0}
\setcounter{figure}{0}

\section*{Supplemental material for ``Spontaneous Twirls and Structural Frustration in Moir\'e Materials''}

\subsection{Relaxation of elastic bilayers}

In the main text, we describe a model with a single elastic layer on a rigid substrate as a substitute  for the real bilayer system with two elastic layers. Here we explicitly show the mapping between the two models. The equilibrium configuration of the bilayer system is governed by the minimization of the bilayer lattice energy $E = \int d^2\rr(U_t + U_b + V)$, where $t$/$b$ stands for the top/bottom layer,
\begin{equation}
    U_i = \frac{\lambda_i}{2} (\nnabla\cdot\ddelta_i)^2 + \frac{\mu_i}{4} \sum_{\alpha,\beta = x,y} (\partial_\alpha\delta_{i\beta} + \partial_\beta\delta_{i\alpha})^2, \quad i = b, t
\end{equation}
is the elastic energy of layer $i$, and $V = V(\dd_0(\rr) + \ddelta_t(\rr) - \ddelta_b(\rr))$.
Since we have assumed that the two layers have the same Lam\'e constant ratio $\lambda_t/\mu_t = \lambda_b/\mu_b$, the linear substitution
\begin{equation}
    \DDelta(\rr) = \frac{\mu_t\ddelta_t(\rr) + \mu_b\ddelta_b(\rr)}{\mu_t + \mu_b}, \quad \ddelta(\rr) = \ddelta_t(\rr) - \ddelta_b(\rr),
\end{equation}
decouples the total energy functional into a ``center-of-mass'' part
\[
    \int d^2\rr \lp( \frac{\Lambda}{2} (\nnabla\cdot\DDelta)^2 + \frac{\rm M}{4} \sum_{\alpha,\beta = x,y} (\partial_\alpha\Delta_\beta + \partial_\beta\Delta_\alpha)^2 \rp)
\]
and a part associated with relative displacement:
\[
    \int d^2\rr \lp( \frac{\lambda_{\rm eff}}{2} (\nnabla\cdot\ddelta)^2 + \frac{\mu_{\rm eff}}{4} (\partial_\alpha \delta_\beta + \partial_\beta\delta_\alpha)^2 \rp) + \int d^2\rr V\bigl( \dd_0(\rr) + \ddelta(\rr) \bigr),
\]
where
\begin{equation}
    \Lambda = \lambda_t + \lambda_b, \quad {\rm M} = \mu_t + \mu_b, \quad \lambda_{\rm eff} = \frac{\lambda_t\lambda_b}{\lambda_t + \lambda_b}, \quad
    \mu_{\rm eff} = \frac{\mu_t\mu_b}{\mu_t + \mu_b}.
    \label{eq:Lame_eff}
\end{equation}
The relaxation behaviors of $\DDelta(\rr)$ and $\ddelta(\rr)$ are thus independent, the former identical to a suspended (unrelaxed) monolayer and the latter exactly described by our model of a monolayer on rigid substrate with the reduced Lam\'e constants $\lambda_{\rm eff}$ and $\mu_{\rm eff}$.

\subsection{Additional results in triangular lattices}

\begin{figure}
    \centering
    \includegraphics[width=\textwidth]{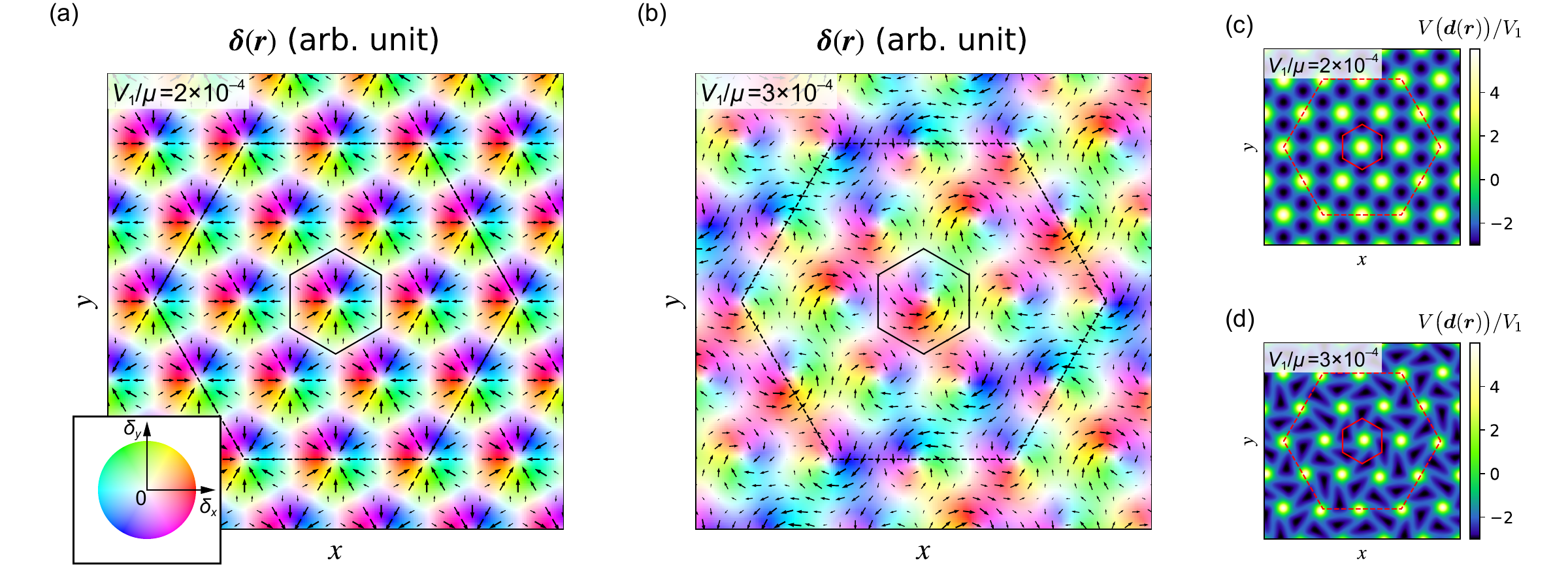}
    \caption{Relaxation results of the triangular moir\'e system with two different $V_1$ values indicated on the top left of each panel, and all other parameters same as shown in Fig. \ref{fig:metastable} in the main text ($\epsilon = -0.017$, $\theta = 0$, $\lambda/\mu = 0.5$ and $\Im V_1 = 0$). (a), (b) The plot of local displacement configuration $\ddelta(\rr)$, represented by both the black arrows and the color pattern with legend at the bottom left of (a). (c), (d) The local stacking energy density $V(\dd(\rr))$ plots, defined in Eq. (\ref{eq:V(d)}), of the two systems.
    }
    \label{figS:varyingV1}
\end{figure}

Fig. \ref{figS:varyingV1} compares the relaxation pattern ($\ddelta(\rr)$ in (a) and (b)) and the relaxed potential pattern ($V(\dd(\rr))$ in (c) and (d)) of the triangular moir\'e system discussed in the main text with a system that has a weaker $V_1$.  This 
comparison explicitly shows that below a critical $V_1$ twirls do not form. 
In this case both the original moir\'e periodicity and the mirror symmetry are preserved.

\subsection{Simulations of the antiferromagnetic lattice $\phi^4$ model}

By defining the dimensionless twirl moments $\tilde\phi_i = \sqrt{2\alpha/\beta}\,\phi_i$, the dimensionless coupling $\tilde J_{ij} = J_{ij}/\beta$ and the dimensionless energy $\tilde E = 2\alpha E/\beta^2$, we can recast the lattice $\phi^4$ model in Eq. (\ref{eq:AFM_phi4}) of the main text to a dimensionless energy functional
\begin{equation}
    \tilde E = \sum_i \lp( \frac{\tilde\phi_i^4}{2} - \tilde\phi_i^2 \rp) + \sum_{i < j} \tilde J_{ij} \tilde\phi_i\tilde\phi_j.
    \label{eqS:AFM_phi4_dimless}
\end{equation}
This reduces the number of free parameters in the model. In the absence of coupling ($\tilde J_{ij} \equiv 0$), the system has exponentially many degenerate ground states where each $\tilde\phi_i$ independently chooses between $\pm 1$.

Now we describe the numerical method we use to minimize $\tilde E$ under periodic boundary condition and the global twist angle constraint (see the discussion around Eq. (\ref{eq:constraint}) in the main text)
\begin{equation}
    \frac{1}{N} \sum_i \tilde\phi_i = \tilde\Phi,
    \label{eqS:constraint}
\end{equation}
where $N$ is the total number of sites in the system. From various random constraint-compliant initial guesses $\tilde\pphi^{(0)} = \lp( \tilde\phi_i^{(0)} \rp)_{i=1}^N$ that will be specified later in the text, we first apply a projected gradient descent method, where in each step we walk the vector $\tilde\pphi$ in the direction opposite to the constraint-projected gradient by some step size $\Delta\tilde\phi$:
\begin{equation}
    \tilde\pphi^{(k+1)} = \tilde\pphi^{(k)} - \frac{\mcP\nabla_{\tilde\pphi}\tilde E(\tilde\pphi^{(k)})}{||\mcP\nabla_{\tilde\pphi}\tilde E(\tilde\pphi^{(k)})||} \cdot \Delta\tilde\phi.
\end{equation}
Here $\mcP$ projects the vector into the $(N-1)$-dimensional hyperplane $\sum_{i=1}^N \tilde\phi_i = 0$. The step size $\Delta\tilde\phi$ is initially set to 1. 
When $\tilde E(\tilde\pphi^{(k+1)}) > \tilde E(\tilde\pphi^{(k)})$ ($\tilde E(\tilde\pphi^{(k+1)}) = \tilde E(\tilde\pphi^{(k)})$), $\Delta\tilde\phi$ is cut in half and the algorithm restarts from $\tilde\pphi^{(k)}$ (continues from $\tilde\pphi^{(k+1)}$). When $\Delta\tilde\phi$ falls below $10^{-6}$, the gradient descent stops and is followed by a multidimensional Newton's method \cite{press2007numerical} aiming at finding the stationary point of the function
\[
    \tilde E'(\tilde\pphi, \tau) = \tilde E(\tilde\pphi) - \tau \lp( \sum_{i=1}^N\tilde\phi_i - N\tilde\Phi \rp),
\]
where $\tau$ is the Lagrange multiplier. Full convergence is declared when all components of $\mcP\nabla_{\tilde\pphi}\tilde E$ fall within $\pm 10^{-14}$.

\begin{figure}
    \centering
    \includegraphics[width=\textwidth]{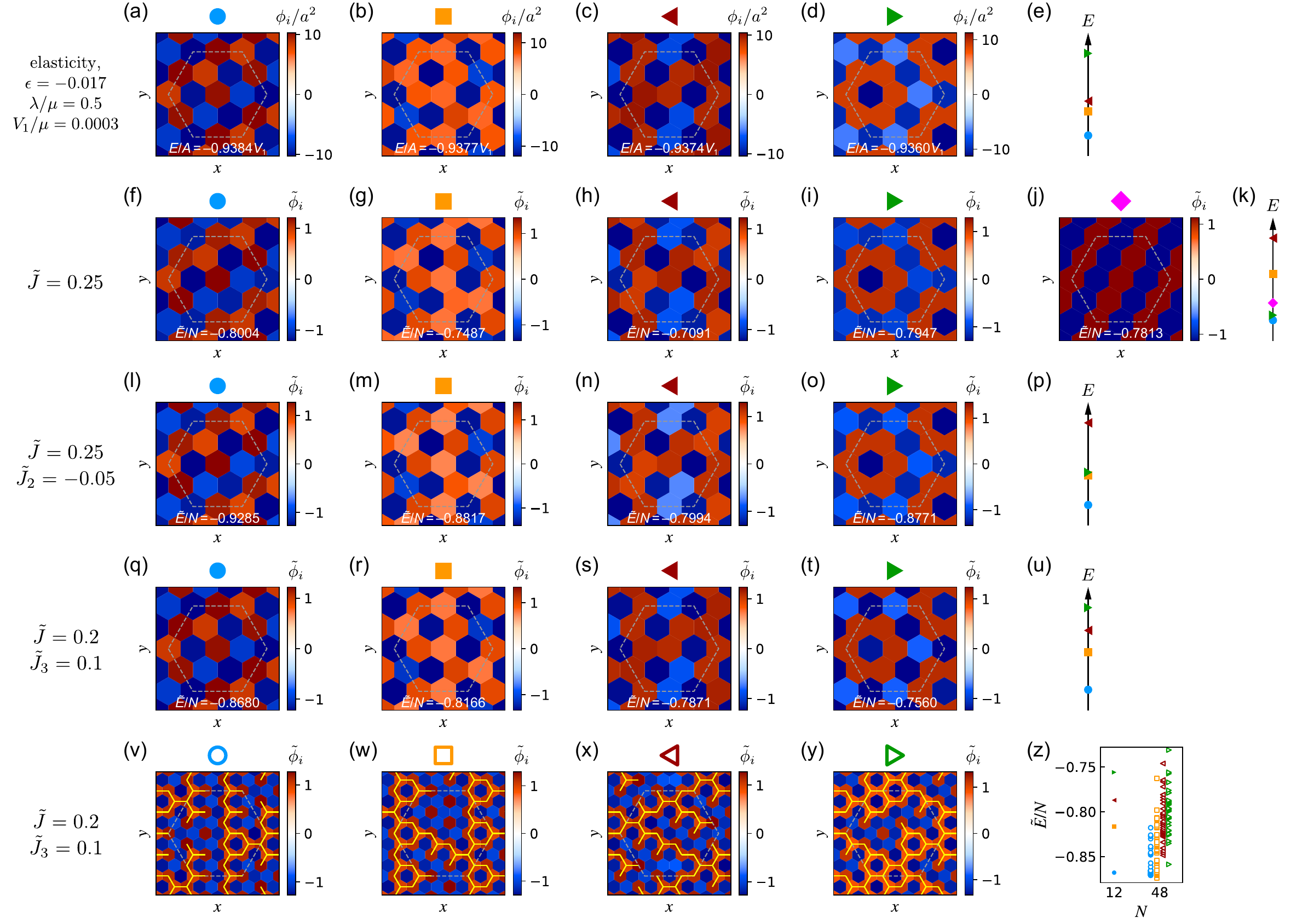}
    \caption{(a)-(d) Twirl moment configurations of the four lowest energy states of the elasticity model. See Figs. \ref{fig:metastable} (a)-(d) in the main text. Total energy density is marked on the bottom of each plot ($A$ is the system area). (e) Relative energies of the 4 states shown in (a)-(d) marked by the symbols on the top of the corresponding panels. (f)-(j) The five lowest energy states of the effective $\phi^4$ model with $2\sqrt{3} \times 2\sqrt{3}$ supercell, where only the nearest-neighbor coupling $\tilde J_{\Mean{ij}} = \tilde J = 0.25$ is included. Rotations, reflections and/or twirl-flips ($\tilde\phi_i \rightarrow -\tilde\phi_i$) have been performed on (f)-(i) to match the orientations of (a)-(d). The plots are ordered based on their similarity to (a)-(d) rather than ascending energy. Average site energy $\tilde E/N$ is marked on the bottom of each plot, where $\tilde E$ is defined in Eq. (\ref{eqS:AFM_phi4_dimless}) and $N$ is the number of sites. (k) Relative energies of the five states shown in (f)-(j). (l)-(u) Similar to (f)-(k), except that further-neighbor couplings are also included. All nonzero coupling constants for each row are shown on the left of each row, with $\tilde J_n$ denoting $n$th-neighbor coupling. (v)-(y) Four representative metastable states of the effective $\phi^4$ model with $4\sqrt{3} \times 4\sqrt{3}$ supercell, obtained by iteration (see detailed algorithm in the text) from the states in (q)-(t) superposed with some noise, respectively. All nearest-neighbor bonds with $\tilde\phi_i > 0$ on both sides are highlighted with yellow lines, from which one can confirm that these four states are not equivalent by symmetry. (z) Comparison of metastable states in $2\sqrt{3} \times 2\sqrt{3}$ ($N = 12$) and $4\sqrt{3} \times 4\sqrt{3}$ ($N = 48$) systems in energy, under $\tilde J = 0.2$ and $\tilde J_3 = 0.1$. Each hollowed data point is obtained by iteration from the (noise-disturbed) $2\sqrt{3} \times 2\sqrt{3}$ state marked with the solid data point with the same color and shape.}
    \label{figS:metastable}
\end{figure}

For $\tilde\Phi = 0$ (corresponding to aligned heterobilayers with $\theta = 0$ described in the main text), we have performed simulations of periodic systems with both $2\sqrt{3} \times 2\sqrt{3}$ and $4\sqrt{3} \times 4\sqrt{3}$ supercells. In the former case, we
use projected unit Gaussian random vectors as initial guesses $\tilde\pphi^{(0)}$, \textit{i.e.} we take $N = 12$ independent standard-normal-distributed random real numbers $\{\tilde\phi_i'\}_{i=1}^{12}$ and let $\tilde\phi_i^{(0)} = \tilde\phi_i' - \sum_{j=1}^N \tilde\phi_j'/N$. Many independent trials are run to ensure that all metastable states are obtained. In the $4\sqrt{3} \times 4\sqrt{3}$ system, the number of metastable states
increases exponentially and we only find a small fraction of the full set of 
metastable states if we start from completely random initial guesses. 
Instead, we start from the four lowest energy states in the $2\sqrt{3} \times 2\sqrt{3}$
subsystems, adding various levels of noise (\textit{i.e.} $\delta\tilde\phi$ times the projected unit Gaussian random vector, where $\delta\tilde\phi$ varies from 0 to 1 in increments of 0.1 and for each nonzero $\delta\tilde\phi$ value we run 10 independent trials).

Some representative results from our $\phi^4$ model simulation are shown in Fig. \ref{figS:metastable} together with the elasticity model results. We see that though the pure nearest-neighbor $\phi^4$ formulation qualitatively reproduces the twirl configurations of the lowest four energy configurations of the elasticity formulation individually, it does not correctly reproduce their energy hierarchy and part of twirl moment hierarchies within individual states. Moreover, it predicts an extra stripy state shown in Fig. \ref{figS:metastable} (j), which is the third lowest in energy in the $\phi^4$ model but is not reported in our elasticity model simulation. Including a small ferromagnetic second-neighbor coupling ($\tilde J_2 < 0$) partially fixes the energy hierarchy and removes the extra stripy state, but does not fix the twirl moment hierarchies. Interestingly, including antiferromagnetic third-neighbor coupling ($\tilde J_3 > 0$) solves all these problems within the first 4 states. We speculate that including both second- and third-neighbor couplings may help to reach better quantitative agreement with the elasticity model results. However, states with higher energies are significantly more difficult to reproduce well. Higher-order couplings involving more than 2 twirls may be necessary in those regimes. Nevertheless, we believe that the essential physics of triangular-lattice frustration is captured by our effective $\phi^4$ model, as we see from Fig. \ref{figS:metastable} (z) that enlarging the supercell to $4\sqrt{3} \times 4\sqrt{3}$ leads to very dense metastable state energy spectra. 

For a rough extraction of model parameters $\alpha$ and $\beta$, we compare the difference in energy between the lowest two energy states in both the original model (Fig. \ref{figS:metastable} (a), (b)) and the effective model (Fig. \ref{figS:metastable} (q), (r)), which gives $\Delta E/A = 0.0007V_1$ and $\Delta\tilde E/N = 0.0514$. According to the conversion between the original and dimensionless energies, this gives $2\alpha/\beta^2 = \Delta\tilde E/\Delta E \approx 73/A_{\rm MC}V_1$ where $A_{\rm MC} = A/N$ is the area of moir\'e cell. We also compare the scales of $\tilde\phi_i$ and $\phi_i$, which gives $\sqrt{2\alpha/\beta} = \tilde\phi_i/\phi_i \approx 1/9a^2 \approx 333/A_{\rm MC}$ (note that $A_{\rm MC} = (\sqrt{3}a^2/2)/\epsilon^2$ where $\epsilon = -0.017$). Combining these two conditions (note that $V_1 = 3\times10^{-4}\mu$), we get $\beta \sim 0.5\mu/A_{\rm MC}$ and $\alpha \sim 2.5\times10^4\mu/A_{\rm MC}^3 \approx 10^{-6} \mu/a^3$.

\subsection{Twist angle dependence in continuum and triangular lattice $\phi^4$ models}

\begin{figure}
    \centering
    \includegraphics[width=\textwidth]{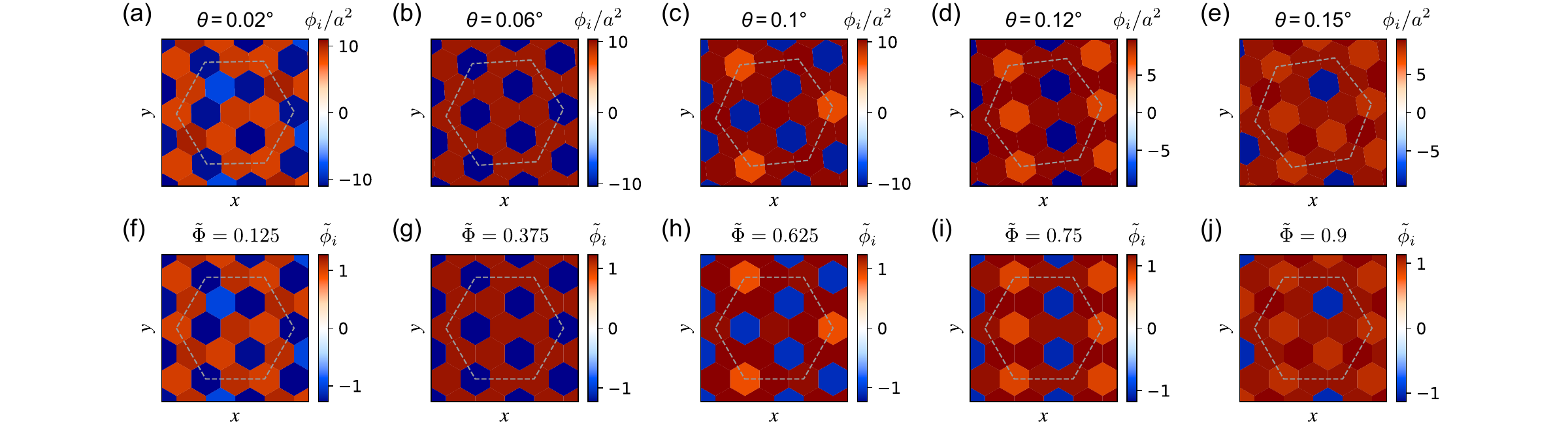}
    \caption{(a)-(e) The twirl moment configurations of the lowest energy configurations of the triangular-system continuum elasticity model with the same parameter as plotted in Fig. \ref{fig:introduction} (a), except that various finite global twist angles $\theta$ are introduced, as indicated on the top of each panel. (f)-(j) Simulated lowest-energy configurations of the effective $\phi^4$ model (Eq. (\ref{eqS:AFM_phi4_dimless})) with $\tilde J = 0.2$, $\tilde J_3 = 0.1$ and various constrained averages $\tilde\Phi$ indicated on the top of each panel. Necessary symmetry transforms including 6-fold rotations, reflections and translations have been applied to some of the plots to reflect smooth evolution with $\theta$ as well as the match between the results of the two models.}
    \label{figS:twist_dependence}
\end{figure}

In Fig. \ref{figS:twist_dependence}, we show the dependence of the triangular-system twirl moment configuration on the global twist angle. For the five twist angles presented in Figs. \ref{figS:twist_dependence} (a)-(e), there are respectively 7, 8, 9, 10 and 11 left twirls ($\tilde\phi_i > 0$) out of 12 twirls in each $2\sqrt{3} \times 2\sqrt{3}$ supercell. Figs. \ref{figS:twist_dependence} (f)-(j) reveal a similar pattern by varying the average twirl moment $\tilde\Phi$ in our effective $\phi^4$ model with $2\sqrt{3} \times 2\sqrt{3}$ superlattice periodicity and the ``best'' set of coupling strengths we have identified: $\tilde J = 0.2$, $\tilde J_3 = 0.1$. We see a proportional relation between $\tilde\Phi$ and $\theta$: $\tilde\Phi \sim 6.25\theta[^\circ]$ (\textit{i.e.} $\tilde\Phi/\Phi \sim 358/A_{\rm MC}$, where $\Phi$ is the dimensionful average twirl moment), which is slightly less accurate at relatively large twist angles. This relation agrees well with the one we have extracted for individual twirls from the last section: $\tilde\phi_i/\phi_i \sim 333/A_{\rm MC}$, justifying the approximation that the coupling strength is independent of the twist angle within the regime of nontrivial twirl configurations.

\subsection{Analytical solution of square-lattice $\phi^4$ model}

The N\'eel antiferromagnetism in our square-lattice effective $\phi^4$ model can be analytically solved. Here we describe the process: apply Eq. (\ref{eqS:AFM_phi4_dimless}) to a large simple of square lattice with $N$ sites and pure nearest-neighbor (dimensionless) coupling strength $\tilde J$, and we get, noting that each site couples with 4 nearest neighbors and that each bond connects 2 sites,
\begin{equation}
    \tilde E(\tilde\phi_1, \tilde\phi_2) = \frac{N}{2} \lp( \frac{\tilde\phi_1^4}{2} - \tilde\phi_1^2 + \frac{\tilde\phi_2^4}{2} - \tilde\phi_2^2\rp) + 2N\tilde J\tilde\phi_1\tilde\phi_2.
\end{equation}
Our goal is to minimize $\tilde E$ under the global twist angle constraint that the twirl moments average  $\tilde\Phi$ (Eq. (\ref{eqS:constraint})) must be proportional to $\sin\theta$. The constrained minimization is realized by introducing Lagrange multiplier $\tau$ and extremizing
\begin{equation}
    \tilde F(\tilde\phi_1, \tilde\phi_2, \tau) = \tilde E(\tilde\phi_1, \tilde\phi_2) - N\tau\lp( \frac{\tilde\phi_1 + \tilde\phi_2}{2} -\tilde\Phi \rp),
\end{equation}
which gives the equation
\begin{equation}
    \tilde\phi_1^3 - \tilde\phi_1 + 2\tilde J\tilde\phi_2 = \tilde\phi_2^3 - \tilde\phi_2 + 2\tilde J\tilde\phi_1 = \frac{\tau}{2}.
\end{equation}
Taking the difference between the left-hand side and the intermediate expression and dividing it by $\tilde\phi_1 - \tilde\phi_2$ (note that $\tilde\phi_1 = \tilde\phi_2$ gives ferromagnetic state rather than antiferromagnetic state), and we get
\begin{equation}
    \tilde\phi_1^2 + \tilde\phi_1\tilde\phi_2 + \tilde\phi_2^2 = 1 + 2\tilde J.
    \label{eqS:analytic}
\end{equation}
Combining Eq. (\ref{eqS:analytic}) with $\tilde\phi_1^2 + 2\tilde\phi_1\tilde\phi_2 + \tilde\phi_2^2 = (\tilde\phi_1 + \tilde\phi_2)^2 = 4\tilde\Phi^2$ gives $(\tilde\phi_1 - \tilde\phi_2)^2 = 4 + 8\tilde J - 12\tilde\Phi^2$, \textit{i.e.}, (assuming $\tilde\phi_1 > \tilde\phi_2$)
\begin{equation}
    \tilde\phi_1 = \tilde\Phi + \sqrt{1+2\tilde J - 3\tilde\Phi^2}, \quad \tilde\phi_2 = \tilde\Phi - \sqrt{1+2\tilde J - 3\tilde\Phi^2}.
\end{equation}
These solutions exist only when $\tilde\Phi$ is below the critical value
\begin{equation}
    \tilde\Phi_c = \sqrt{\frac{1+2\tilde J}{3}}.
\end{equation}
Otherwise, our assumption that $\tilde\phi_1 \ne \tilde\phi_2$ fails and only the ferromagnetic solution $\tilde\phi_1 = \tilde\phi_2 = \tilde\Phi$ is left. Comparing the energies of the antiferromagnetic and ferromagnetic solutions, we find that the antiferromagnetic solution is preferred whenever it exists. Fig. \ref{figE:square1} (e) in the End Matter is plotted based on this analytic solution, with the model parameters converted to their original dimensionful forms (note that $\phi_i/\sqrt{(\beta + 2J) / 2\alpha} = \tilde\phi_i/\sqrt{1 + 2\tilde J}$ and $\Phi/\sqrt{(\beta + 2J) / 2\alpha} = \tilde\Phi/\sqrt{1 + 2\tilde J}$).

Unlike the triangular case, here we cannot extract the values of $\alpha$, $\beta$ and $J$ by comparing Figs. \ref{figE:square1} (d) and (e) because the shapes of the curves in (e) are fixed, giving no further conditions than the two equations from overall scalings of energies and twirl moments. The parameters would have to be determined from analysis of the microscopic mechanism. However, we note that $\tilde J$ must be large enough to ensure that the simple $\sqrt{2} \times \sqrt{2}$ antiferromagnetic N\'eel state is favored over the states with longer-period translational symmetry breaking.
Indeed, from Eq (\ref{eqS:AFM_phi4_dimless}), for small $\tilde J_{ij}$, the twirls are only weakly interacting with $|\tilde\phi_i| \approx 1$; therefore, to obtain a small average twist $\tilde\Phi \approx (N_{\rm left} - N_{\rm right})/(N_{\rm left} + N_{\rm right})$, one needs to have a large and unequal number of left and right twirls within a supercell.

\subsection{Hysteresis of the antiferromganetic triangular-lattice $\phi^4$ model}

\begin{figure}
    \centering
    \includegraphics[width=\textwidth]{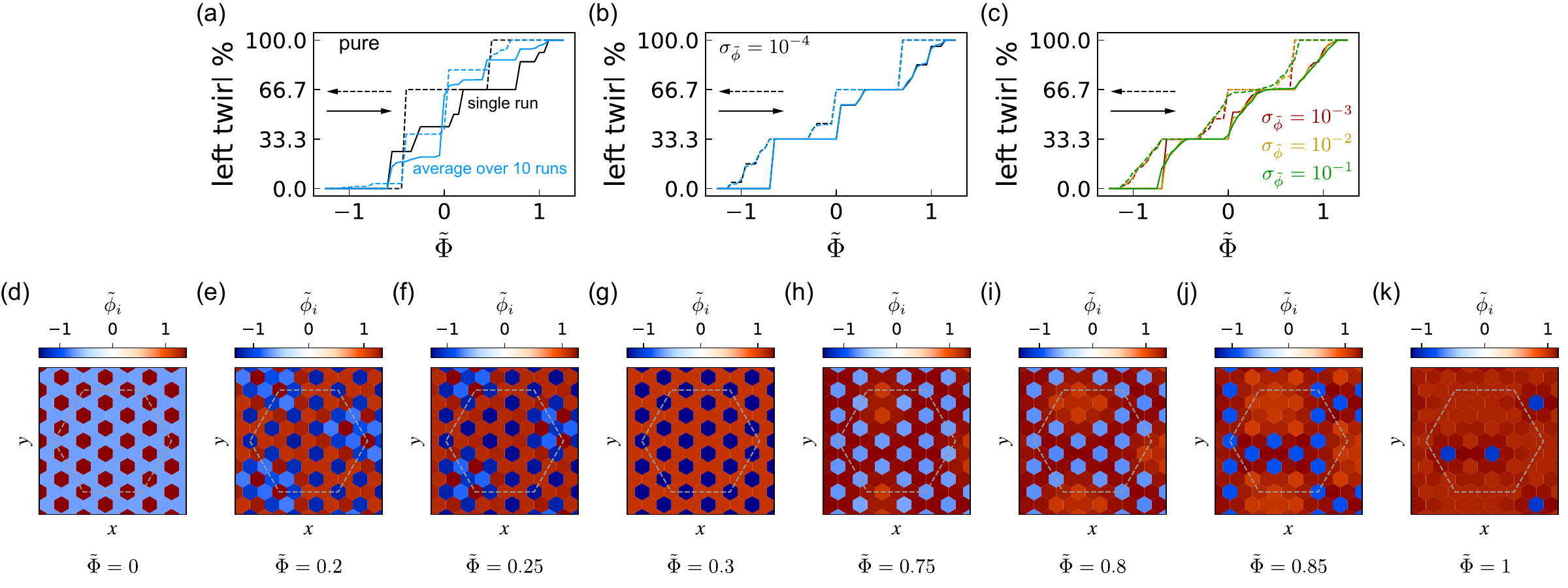}
    \caption{(a)-(b) Evolution of the left twirl percentage in the $4\sqrt{3} \times 4\sqrt{3}$ periodic triangular-lattice $\phi^4$ model with $\tilde J = 0.2$ and $\tilde J_3 = 0.1$, along a loop of $\tilde\Phi$ between $-1.25$ and $1.25$ in increments of $\pm 0.05$, from both (a) pure runs and (b) noisy runs with $\sigma_{\tilde\phi} = 10^{-4}$ (see definition in the text). Black and blue lines correspond to a specific run and the average of 10 independent runs. In (b), the black line is almost covered by the blue line. Forward and backward sweeps are respectively represented by solid and dashed lines. (c) Evolution of the left twirl percentage under 3 different noise levels, $\sigma_{\tilde\phi} = 10^{-3}$, $10^{-2}$ and $10^{-1}$, indicated by 3 colors. Each curve represents the average result of 10 independent runs. (d)-(k) Snapshots of the twirl configuration evolution along the forward sweep in the single run of (b), with the $\tilde\Phi$ value labeled on the bottom of each panel.}
    \label{figS:hysteresis}
\end{figure}

Similar to what we have done for the elasticity model in the main text, here we perform serial simulations of our triangular-lattice antiferromagnetic $\phi^4$ model with the enlarged $4\sqrt{3} \times 4\sqrt{3}$ supercell, along a loop of $\tilde\Phi$ from $-1.25$ to $1.25$ and back in increments of $\Delta\tilde\Phi = \pm0.05$. The initial sample ($\tilde\Phi = -1.25$) is obtained by iterating from a random constraint-compliant initial guess $\tilde\pphi^{(0)}$, as we have described before. The initial states in subsequent samples are taken from the result of the previous run with a necessary adjustment $\Delta\tilde\phi_i = \Delta\tilde\Phi$ reflecting the change in $\tilde\Phi$. Besides this, we also perform serial simulations of the same system with artificially introduced noise. In particular, at each inter-sample transition, besides the adjustment $\Delta\tilde\Phi$, we also add an extra projected Gaussian noise $\delta\tilde\phi_i = \delta\tilde\phi_i' - \sum_{j=1}^N \delta\tilde\phi_j'/N$, where the $\delta\tilde\phi_i'$'s are site-independent Gaussian random variables with standard deviation $\sigma_{\tilde\phi}$. We refer to simulations with $\sigma_{\tilde\phi} = 0$ and $\sigma_{\tilde\phi} \ne 0$ respectively as ``pure'' and ``noisy'' runs. At each noise level $\sigma_{\tilde\phi}$ we independently run 10 loops, and the results are summarized in Figs. \ref{figS:hysteresis} (a)-(c).

Like in the main text, hysteretic behaviors in the left twirl ratio (the number ratio of sites with $\tilde\phi_i > 0$) are observed in both pure and noisy runs. However, as Fig. \ref{figS:hysteresis} (a) shows, the result of the pure simulation strongly varies across our 10 independent runs, even though all of them start from the uniformly aligned configuration ($\tilde\Phi = -1.25$, left twirl ratio $=0$). This ``chaotic'' behavior is consistent with the existence of exponentially many metastable states in the intermediate region of $\tilde\Phi$ due to the large system size compared to the $2\sqrt{3} \times 2\sqrt{3}$ system studied in the main text. Strikingly, even a weak noise ($\sigma_{\tilde\phi} = 10^{-4}$) can help to eliminate this ``chaos" and stabilize the plateaus at $1/3$ and $2/3$, as shown in Fig. \ref{figS:hysteresis} (b).
In addition, Fig. \ref{figS:hysteresis} (c) shows that higher levels of noise tend to smoothen the evolution of twirl configurations. These phenomena can be understood by noting that in each sample, the role of noise is partly to select the state with the lowest energy from all the metastable states in the vicinity of the previous sample in the configuration space $\{\tilde\phi_i\}$.

In Figs. \ref{figS:hysteresis} (d)-(k), we show typical evolution of twirl configurations with $\tilde\Phi$. We see Kekul\'e-like $\sqrt{3} \times \sqrt{3}$ patterns in the $1/3$ (d) and $2/3$ (g) plateaus. Between the two plateaus, we see clusters of Kekul\'e-like phases whose sizes vary with $\tilde\Phi$; Above the $2/3$ plateau, clusters of aligned (``ferromagnetic'') phase appear and expand. These behaviors are akin to the magnetic domain expansion and shrinking in ferromagnets in response to change in magnetic field, except that the whole process is copied three times with some of the saturated ferromagnetic states replaced by Kekul\'e-like states. We expect that 3-step hysteresis between Kekul\'e plateaus can also be displayed in response to the ``external magnetic field'' in the lattice $\phi^4$ model. Similar physics was discussed in triangular-lattice Ising antiferromagnets \cite{kageyama1997fieldinduced, kudasov2006steplike}, 
though in our model, due to the softness of ``spins" it is possible to vary ``magnetization" without changing the twirl chirality configuration.

\subsection{Effect of perpendicular electric field in triangular systems}

\begin{figure}
	\centering
	\includegraphics[width=\textwidth]{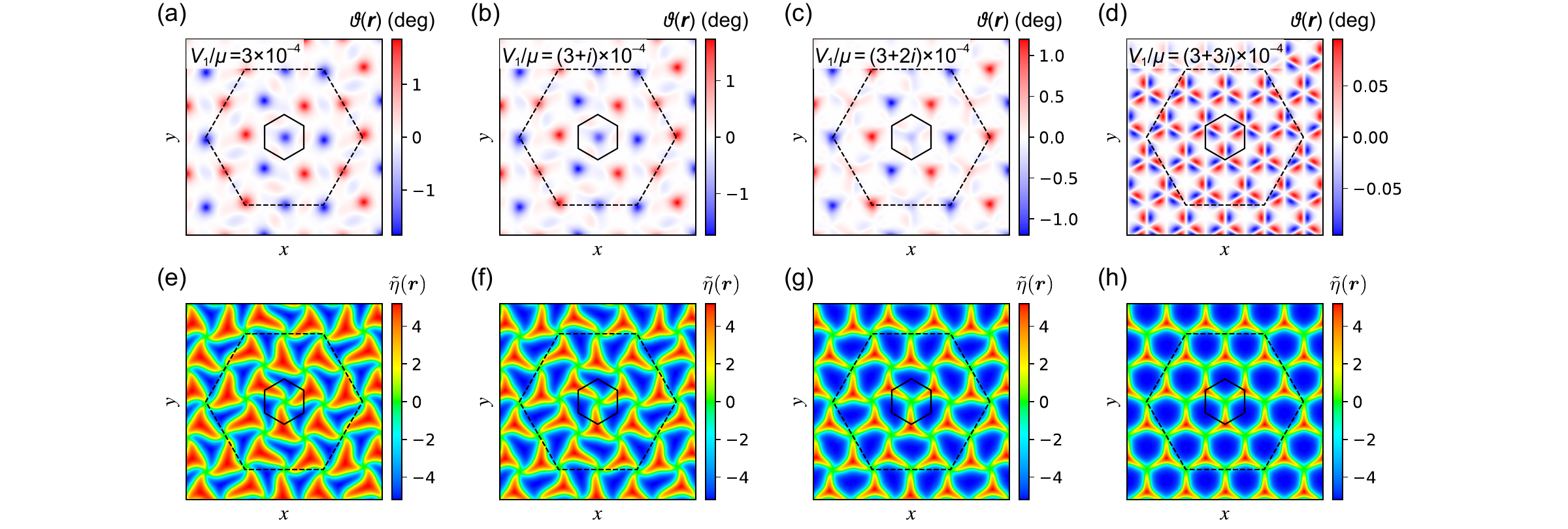}
	\caption{(a)-(d) Local twist angle plots of the lowest-energy configuration of the triangular system with the same parameters as Fig. \ref{fig:metastable} in the main text, except that $V_1$ acquires an imaginary part, as indicated on the top left of each panel. (e)-(h) Plots of $\tilde\eta(\rr)$ defined by Eq. (\ref{eqS:eta}), for the 4 configurations corresponding to (a)-(d).
    For each of the 4 presented configurations, we have selected the one from many symmetry-related lowest-energy states we obtain from convergent iteration, so that they show smooth evolution with $\Im V_1$.}
	\label{figS:ImV1}
\end{figure}

In real systems like heterobilayer TMDs, an imaginary part of $V_1$ due to inequivalent honeycomb sublattices is inevitable, and is often widely tunable with an out-of-plane electric field.
Fig. \ref{figS:ImV1} (a)-(d) shows the evolution of the local twist angle configuration of the lowest-energy state as $\Im V_1$ increases. We see that while the frustrated pattern in the twirl configurations is preserved at small $\Im V_1$ [Fig. \ref{figS:ImV1} (b)], some twirls shrink faster than others with $\Im V_1$. In particular, Fig. \ref{figS:ImV1} (c) shows a residue of Kekul\'e order in twirl configuration at intermediate $\Im V_1$. When $\Im V_1$ approaches $\Re V_1$, the spontaneous twirls disappear, only leaving a background pattern of small local twist angle that respects the original moir\'e periodicity as shown in Fig. \ref{figS:ImV1} (d). 

We also introduce the indicator function
\begin{equation}
    \tilde\eta(\rr) = \eta(\dd(\rr)) = i\sum_{j=0}^5 (-1)^{j}e^{i\GG_j\cdot\dd(\rr)},
    \label{eqS:eta}
\end{equation}
which changes sign at the domain walls between metastable stacking regions, to examine the behavior of domain structures.
$\tilde\eta(\rr)$ is plotted in Figs. \ref{figS:ImV1} (e)-(h) for the four configurations.
As $\Im V_1$ increases, in addition to the previously reported unbalancing between upward-triangle (red) and downward-triangle (blue) domains \cite{enaldiev2022scalable, bennett2022theory, ramos-alonso2025flat}, we also see a gradual reduction in translational-symmetry-breaking distortion in the domain wall network that is associated with the spontaneous twirls.

Since all the metastable states are close in energy, in a physical process, as $\Im V_1$ decreases from maximum, we expect the system to randomly choose a symmetry-broken state with some frustrated configuration of twirl chiralities, which will later freeze. With the assistance from external disorder, defects or impurities, certain short-range orders may be stabilized.
Electric-field-controlled crystal-glass transition in moir\'e materials hereby may be potentially realized. Possible description of twirl behaviors by effective lattice models in $\Im V_1 \ne 0$ cases will be considered in future study.

\subsection{Estimated Model Parameters in Real Triangular Systems}

In non-centrosymmetric bilayers the stacking energy
\begin{equation}
    V(\dd) = V^0(\dd) - \mcE P(\dd) + O(\mcE^2)
\end{equation}
consists of the adhesion part $V^0(\dd)$ and the electrostatic part that depends on the applied perpendicular electric field $\mcE$. $P(\dd)$ is the electric polarization, which arises from the difference in electronegativity between aligned atoms in the two layers. We note that a phase shift in the complex parameter $V_1 \rightarrow V_1 e^{\pm 2\pi i/3}$ can be compensated by a redefinition of the stacking parameter $\dd(\rr) \rightarrow \dd(\rr) \pm \ttau_0$, where $\ttau_0 = (0,\, a/\sqrt{3})$ is the honeycomb's nearest-neighbor bond vector.

\begin{table}[h]
    \centering
    \setlength{\tabcolsep}{5pt}
    \begin{tabular}{|c|c|c|c|c|}
        \hline
        system & $\epsilon$ & $V_1^0$ (meV/\AA$^2$) & $\gamma^0 = \Re V_1^0/\mu_{\rm eff}\epsilon^2$ & $P_1$ ($e$/$\upmu$m)
        \\\hline
        parallel \ce{MoS_2}/\ce{MoSe_2} & \multirow{2}{*}{$-0.041$\cite{haastrup2018computational, gjerding2021recent}} & $1.23-0.0737i$ & 0.41 & $-53.9+53.0i$
        \\\cline{1-1} \cline{3-5}
        antiparallel \ce{MoS_2}/\ce{MoSe_2} & & $1.06+0.192i$ & 0.35 & $-47.8-1.04i$
        \\\hline
        parallel \ce{WS_2}/\ce{WSe_2} & \multirow{2}{*}{$-0.040$\cite{haastrup2018computational, gjerding2021recent}} & $1.22-0.104i$ & 0.42 & $-44.9+55.4i$
        \\\cline{1-1} \cline{3-5} antiparallel \ce{WS_2}/\ce{WSe_2} & & $1.05+0.229i$ & 0.33 & $-41.0-4.13i$
        \\\hline
        parallel \ce{MoS_2}/\ce{WS_2} & $0.002$\cite{kaliteevski2023twirling} & $1.37+0.0i$ & 190 & $-17.8+66.9i$
        \\\hline
        parallel \ce{MoSe_2}/\ce{WSe_2} & 0.004\cite{kaliteevski2023twirling} & $1.08+0.0i$ & 37.5 & $-9.89+42.7i$
        \\\hline
    \end{tabular}
    \caption{A compilation of first-principle-based model parameter values for various heterobilayer systems. The lattice constant mismatch $\epsilon$ in
    \ce{MoS_2}/\ce{MoSe_2} and \ce{WS_2}/\ce{WSe_2} is based on the lattice constant values provided by Computational 2D Materials Database \cite{haastrup2018computational, gjerding2021recent}, and those of \ce{MoS_2}/\ce{WS_2} and \ce{MoSe_2}/\ce{WSe_2} are directly given by Ref. \cite{kaliteevski2023twirling}.
    $V_1^0$ and $P_1$, the first-star Fourier coefficients of $V^0(\dd)$ and $P(\dd)$, are obtained from  Ref. \cite{bennett2022theory} with a scaling correction explained in Ref. \cite{ramos-alonso2025flat}. For antiparallel bilayers we have applied the phase $e^{2\pi i/3}$ described in the text to rotate $V_1^0$ closer to the real axis and also rotate $P_1$ accordingly. The bilayer reduced Lam\'e coefficient is taken as $\mu_{\rm eff} = \mu_{\ce{MoS_2}}/2 = 1.8\,{\rm eV/\text{\AA}^2}$ \cite{carr2018relaxation}, assuming that all TMD materials have the same Lam\'e constant.}
    \label{table:modelparams}
\end{table}

The estimated parameter values of some TMD heterobilayers are summarized in Table \ref{table:modelparams}, where $V^0(\dd)$ and $P(\dd)$ are characterized by their first-star Fourier coefficients $V_1^0$ and $P_1$.
Here we explain the estimation. From the Supplementary Information of Ref. \cite{bennett2022theory} we get the list of \textit{ab initio}-based values of $V_1^0A_c$ and $P_1A_c$ for TMD systems (note an extra factor 2 between Ref. \cite{bennett2022theory} and our convention), which are then divided by the unit cell area $A_c = \sqrt{3}a^2/2$ to obtain $V_1^0$ and $P_1$. Here $a$ is the average lattice constant of the two materials. Furthermore, following the description in the Supplemental Information of Ref. \cite{ramos-alonso2025flat}, we upscale the $P_1$ values by a factor of 89 to account for the difference between numerical predictions and experimental results, which may be due to quantum fluctuation and additional sources of disorder \cite{tang2023sliding}. The shift in convention $\lp( V_1^0, P_1 \rp) \rightarrow \lp( V_1^0 e^{\pm2\pi i/3}, P_1 e^{\pm2\pi i/3} \rp)$ discussed above is also applied to some systems to move $V_1^0$ closer to the real axis.

We define the dimensionless parameter $\gamma = \Re V_1/\mu_{\rm eff}\epsilon^2$ and show its value under zero electric field for various systems in Table \ref{table:modelparams}. As argued in the main text, $\gamma$ needs to be at least of order 1 to form spontaneous twirls. In parallel TMD heterobilayers with the same type of chalcogen atoms, $\gamma$ is significantly larger than 1 due to small lattice mismatch $\epsilon$, favoring spontaneous twirls in aligned bilayers; In TMD heterobilayers with the same type of metal atoms, $\gamma$ is typically only a fraction of 1 due to relatively large $|\epsilon|$. However, $\gamma$ can be increased by applying an electric field $\mcE > 0$, which gives $V_1 \approx V_1^0 - \mcE P_1$. For example, applying $\mcE = 100\,\rm mV/\AA$ to parallel \ce{MoS_2}/\ce{MoSe_2} raises $\gamma$ from $0.41$ to about $0.6$. In parallel systems, the electric field also has the side effect of raising $|\Im V_1|$, which is unfavorable to twirling and frustration. We note, however, that for some antiparallel systems this side effect is minimal due to relatively small imaginary part of $P_1$.

\end{document}